\documentclass[12pt]{article}
\usepackage{amsmath}
\usepackage{amssymb}
\usepackage{cite}

\def\bq{\begin{equation}}
\def\ee{\end{equation}}
\def\bea{\begin{eqnarray}}
\def\eea{\end{eqnarray}}

\usepackage{epsfig}
\def\FIGURE#1{\begin{figure}[ht]#1\end{figure}}

\makeatletter

\begin{document}
\begin{titlepage}
\thispagestyle{empty}

\bigskip

\begin{center}
\noindent{\Large \textbf
{A Transfer Matrix Method for Resonances in Randall-Sundrum Models II: The Deformed Case}}\\

\vspace{0,5cm}

\noindent{R. R. Landim ${}^{a}$, G. Alencar ${}^{b}$\footnote{e-mail: geovamaciel@gmail.com },M. O. Tahim ${}^{b}$ and R.N. Costa Filho ${}^{a}$}

\vspace{0,5cm}

 {\it ${}^a$Departamento de F\'{\i}sica, Universidade Federal do Cear\'{a}-
Caixa Postal 6030, Campus do Pici, 60455-760, Fortaleza, Cear\'{a}, Brazil. 
 }

\vspace{0.2cm}

{\it ${}^b$Universidade Estadual do Cear\'a, Faculdade de Educa\c c\~ao, Ci\^encias e Letras do Sert\~ao Central- 
R. Epitcio Pessoa, 2554, 63.900-000  Quixad\'{a}, Cear\'{a},  Brazil.}

\end{center}

\vspace{0.3cm}

\begin{abstract}

Here we consider resonances of the Gauge, Gravity and Spinor fields in Randall-Sundrum-like scenarios. We consider membranes that are generated by a class of topological defects that are deformed  domain walls obtained from other previously known ones. They are obtained by a deformation procedure generate different potentials to the  associated Schr\"odinger-like equation. The resonance spectra are calculated numerically using the method of Transfer Matrix developed by the authors and presented in JHEP 1108 (2011) 071. The new deformed defects change the resonances spectra of all fields considered and the associated phenomenology as well.

\end{abstract}


\end{titlepage}

\section{Introduction}

The core idea of extra dimensional models is to consider the four-dimensional
universe as a hyper-surface embedded in a multidimensional manifold. This original proposal of Kaluza and Klein \cite{Appelquist:1988fh} did not attracted much attention at the time. However, this situation changed drastically nowadays after the advent of supergravity and superstring theory where the extra dimensions models are a necessary ingredient \cite{Polchinski:1998rq,Polchinski:1998rr,Berkovits:2000fe}. More specifically, after Randall and Sundrum's idea of a Brane world with non-factorisable metric,
there have been an extensive use of these ideas \cite{Randall:1999vf,Randall:1999ee}.  
This model provides a possible solution to the hierarchy problem and show how gravity is trapped to a membrane. 
It has been found that the trapping of fields and the associated effective action depends on an associated equation that depends only on the extra dimension $y$ \cite{Kehagias:2000au}. An interesting aspect of these models is that the associated equation can be written in a  form very similar to the Schr\"odinger equation with an associated potential. Therefore, the hierarchy problem is reduced to the  resolution of a one dimensional quantum mechanics problem. That enriches the problem since
many known tools can be used to investigate the physics of the problem.

An important example of one dimensional quantum mechanics is the well known result of electrons moving in semiconductor heterostructures.  In such structures, each material composing the heterostructure are considered potential barriers felted by the electrons. By solving the Schr\"odinger equation we can compute transmission and reflection coefficients, making possible the understanding of the transport properties and the resonance structures of structure defined by a one-dimensional potential. From the theoretical point of view such systems have been extensively studied  in the 90's \cite{kim:1988,renan:1993,renan:1994}. In particular, because of the improvement of  numerical computational methods and computer processors,  a few more difficult problems without exact solutions were attacked. The source of those difficulties is the complexity of the several kinds of potentials 
that should be treated in the Schr\"odinger equation. As an example, a numerical method was developed in order to solve the problem of potential barriers of any shape. In order 
to obtain these results numerically, they made use of a common tool in condensed matter systems called the transfer matrix method \cite{ando:1987}.

The main difference between the Randall-Sundrum(RS) and Kaluza-Klein models is that the later has  a compact extra dimension and the former  has an infinity one. As consequence the  Kaluza-Klein models have a discrete spectrum of mass that depends on the radius of the extra dimension. In this scenario the spectrum in four dimension is trivially recovered. 
In RS models the extra dimension is infinity and the mass spectrum is determined by a Schr\"odinger like 
equation where the mass spectrum may not be discrete. In both models it is important to verify if the  electromagnetism and gravity are reproduced in first approximation inside the brane. That happens only if the zero mode, 
or massless mode, is localized in the membrane. In these scenarios it has been shown that the zero mode of the gravity field and the left hand fermion are localized\cite{Kehagias:2000au}. 

Another class of fields are the gauge or form fields. The presence of one more extra dimension ($D=5$) provides the existence of many antisymmetric fields, namely the $0,1,2,3,4$ and $5-$forms. However, gauge freedom can be used 
to cancel the dynamics of the $4$ and $5-$form fields in the visible brane. These fields has been used to describe space-time torsion and the axion \cite{Mukhopadhyaya:2004cc,Arvanitaki:2009fg,Svrcek:2006yi} that 
have separated descriptions by the two-form. Besides this, String Theory shows the naturalness of higher rank tensor fields in its spectrum \cite{Polchinski:1998rq,Polchinski:1998rr}. Other applications of these kind
of fields have been made showing its relation with the AdS/CFT conjecture \cite{Germani:2004jf}. However, it has been shown that only the $0-$form is localized \cite{Kehagias:2000au}. The mass spectrum of the two and three form and the respective coupling with the dilaton have been studied, for example, in Refs. \cite{Mukhopadhyaya:2007jn,DeRisi:2007dn,Mukhopadhyaya:2009gp,Alencar:2010mi}. 
The study of this kind of coupling, inspired in string theory, is important in order to produce a process that, in principle, could be seen in LHC. This is a Drell-Yang process in which a pair quark-antiquark 
can give rise to a three(two)-form field, mediated by a dilaton. 

The RS model has also been modified to consider the membrane as a topological defect generated by a scalar field \cite{Bazeia:2008zx}. In this approach, the soliton-like solutions are used to simulate the membrane, and several kinds of defects in brane scenarios are considered in the literature \cite{Gremm:1999pj,Brihaye:2010nf,yves:b,yves:c}. This models can be generalized to obtain a rich class of defects by a deformation procedure of the $\lambda\pi^4$ potential \cite{Bazeia:2002xg}. It is possible to solve the equations of motion by the super-potential method. This formalism was initially introduced in studies about non-super-symmetric domain walls in various dimensions by \cite{DeWolfe:1999cp,Skenderis:1999mm}. As an example, in a recent paper, a model is considered  for coupling fermions to brane and/or antibrane modeled by a kink anti-kink  system \cite{yves:d}. The localization of fields in a framework that considers the brane as a kink have been studied for example in 
\cite{Bazeia:2007nd,Bazeia:2004yw,Bazeia:2003aw,Christiansen:2010aj,Alencar:2010hs,Landim:2010pq,Alencar:2010vk,Fonseca:2011ep}. In many of these works the coupling between the dilaton and the form fields is used to produce the desired localization\cite{Kehagias:2000au}.

In all above-mentioned models the massive spectrum is determined by a Schr\"odinger like equation with a potential that falls to zero at infinity. The spectrum is not discrete and we have a ill defined effective action.  Despite of that, just like in the case of semiconductor heterostructures, there is the possibility of appearance of resonances. This analysis have been done extensively in the literature 
\cite{Bazeia:2005hu,Liu:2009ve,Zhao:2009ja,Liang:2009zzf,Zhao:2010mk,Zhao:2011hg,Li:2010dy,Guo:2011wr,Liu:2011wi,Liu:2009mga,Castro:2010au,Correa:2010zg,Castro:2010uj,Chumbes:2010xg,Castro:2011pp}. 

In order to analyze resonances, we must compute transmission coefficients ($T$), which gives a clear physical interpretation about what happens to a free wave interacting with the membrane. The idea of the existence of a resonant mode is that for a given mass the transmitted and reflected oscillatory modes are in phase inside the membrane, i. e., the transmission coefficient has a peak at this mass value. That means the amplitude of the wave-function has a maximum value at $z=0$ and the probability to find this KK mode inside 
the membrane is higher. This method has been used previously to analyze resonant modes of the gravity, fermion and form fields in a previous work\cite{Landim:2011ki}.

As far as we know there is no study in the literature of resonances for form fields without the dilaton and for gravity and fermion fields with the dilaton in a deformed background. The goal of this piece of work is to apply a 
numerical method to study aspects of the form fields in a scenario of extra dimensions.  We compute the transmission coefficients of Gravity, Fermion and form fields in both setups: with and without the dilaton coupling
in a deformed scenario.  The paper is organized as follows. Section two  is devoted to discuss a solution of the Einstein`s equation with source given by a deformed kink with and without the presence of the dilaton. 
In section three the general prescription to reach the Schr\"odinger equation to the problem is given. We also present in detail the numerical steps to compute the transmission coefficients with transfer matrix.  
In Section four we analyze resonances of the gravity field. In section five, six and seven, we present the resonance structure for $0$,$1$ and $2$ forms respectively. In section eight we present the same study for the 
fermion fields. At the end, we discuss the conclusions and perspectives.

\section{The Deformed Kink as a Membrane}

We start our analysis by studying the space-time background. It is well known that vector gauge fields in these kind of scenarios are not localizable: in four dimensions the gauge vector field theory is conformal and all information coming from warp factors drops out necessarily rendering a non-normalizable four dimensional effective action. However, in the work of Kehagias and Tamvakis \cite{Kehagias:2000au}, it is shown that the coupling between the dilaton and the vector gauge field produces localization of the later. On the other hand, scalar and fermion fields do not need this coupling in order to produce localized zero modes. Therefore, we must consider in this section both backgrounds: with and without the dilaton coupling. 

\subsection{The Background with the Dilaton Coupling}

We start our analysis studying the space-time background. Before analyzing the resonances of fields, it is necessary to obtain
a solution of the equations of motion for the coupled gravity-dilaton-brane system. In a previous manuscript the authors considered two cases: with and without the dilaton coupling. Here we find the solution for the deformed case. As this solution has been studied extensively in the literature we only give a short review of how to reach that. For the case with the dilaton the action is similar given by \cite{Kehagias:2000au}:

\begin{equation}
S=\int d^{5}x \sqrt{-g}\left[2M^{3}R-\frac{1}{2}(\partial\phi)^{2}-\frac{1}{2}%
(\partial\pi)^{2}-V(\phi,\pi)\right],
\end{equation}
where $D=5$, $g$ is the metric determinant and $R$ is the Ricci scalar. Note that we are
working with a model containing two real scalar fields. The field $\phi$ plays the role of
membrane generator of the model while the field $\pi$ represents the dilaton. The
potential function depends on both scalar fields. It is assumed the following ansatz for
the spacetime metric: 
$$ds^{2}=g_{MN}dx^Mdx^N=e^{2A_s(y)}\eta_{\mu\nu}dx^{\mu}dx^{\nu}+e^{2B_s(y)}dy^{2},$$
where $\eta_{\mu\nu}=diag(-1,1,1,1)$ is the metric of the brane, $y$ is the
co-dimension coordinate, and $s$ is a deformation parameter. The deformation parameter
controls the kind of deformed topological defect we want in order to mimic different
classes of membranes. The deformation method is based in deformations of the potential of
models containing solitons in order to produce new and unexpected solutions
\cite{Bazeia:2002xg}. As usual, capital Latin index represents the coordinates in the bulk
and Greek index the ones in the brane. In order to solve this system, we use the so-called superpotential function 
$W_s(\phi)$, defined by 
$$\phi^{\prime}=\frac{\partial W_s}{\partial\phi},$$
following the approach of Kehagias and Tamvakis \cite{Kehagias:2000au}. The
particular solution follows from choosing the potential 
$$
V_s=\exp{\left(\frac{\pi}{\sqrt{12M^{3}}}\right)}\left[\frac{1}{2}\left(\frac{\partial W_s}{%
\partial\phi}\right)^{2}-\frac{5}{32M^{3}}W_s(\phi)^{2}\right],
$$
and superpotential
$$
W_s(\phi)=a\phi^2\left[\frac{s}{2s-1}\left(\frac{v}{\phi}\right)^{1/s}-\frac{s}{2s+1}\left(\frac{\phi}{v}\right)^{1/s}\right],
$$
where $a$ and $v$ are parameters to adjust the dimensionality. As pointed in \cite{Kehagias:2000au}, this potential 
gives us the desired soliton-like solution. In this way, it is easy to
obtain first order differential equations whose solutions are solutions of the
equations of motion above, namely
\begin{equation}
\pi_s=-\sqrt{3M^{3}}A_s,\quad B_s=\frac{A_s}{4},\quad A_s^{\prime}=-\frac{W_s}{12M^{3}}.\label{piBA}
\end{equation}
The solutions for these new set of equations are the following:
\begin{equation}
\phi(y)=v\tanh(ay),  \label{dilat}
\end{equation}
\begin{equation}
A(y)=-\beta_1\left(4\ln\cosh(ay) +\tanh^2(ay)\right) \label{dilat2}
\end{equation}
for $s=1$ and 
\begin{equation}
\phi(y)=v\tanh^s(ay/s),  \label{dilat1}
\end{equation}
\begin{eqnarray}
A_s(y)=-\beta_s\tanh^{2s}\left(\frac{ay}{s}\right)
- \frac{4s}{(2s-1)}\beta_s \biggl{\{}\ln\biggl[\cosh\left(\frac{ay}{s}\right)\biggr]-
\sum_{n=1}^{s-1}\frac1{2n}\tanh^{2n}\left(\frac{ay}{s}\right)\biggr{\}} \label{a}
\end{eqnarray}
for $s>1$, where $\beta_1=\frac{v^2}{72M^3}$ and $\beta_s=\frac{v^2}{12M^3}\frac{s}{(2s+1)}$. 
We should point here that we gave the solution for $s=1$ for completeness, but we are not going to analyze this case since this has been done previously. 

\subsection{The Background without the Dilaton Coupling}

Here we look for the background solution without the dilaton coupling. We could repeat all the steps of the last section to arrive at the final solution. We can also take a short cut and 
from our previous solution we obtain the desired result. By just setting $B_s=\pi=0$ we can get the final answer. The metric is now
\begin{equation}
ds^{2}=e^{2A_s(y)}\eta_{\mu\nu}dx^{\mu}dx^{\nu}+dy^{2},
\end{equation}
with the resulting potential
\begin{equation}
V=\{\frac{1}{2}(\frac{\partial W}{%
\partial\phi})^{2}-\frac{5}{32M^{3}}W(\phi)^{2}\}.
\end{equation}
This solution agrees with the one obtained previously. We must note that the solution for $A_s$ is left unchanged and therefore the contribution of the dilaton is to change the equations of motion. In the next section we present the general set to be used throughout this paper in order to look for the spectrum of field resonances. 

In the limit where $y\rightarrow\infty$ we have for $s=1$:
\begin{equation}
e^{2A(y)}=\frac{e^{-2\beta_1 \tanh^2(ay)}}{(\cosh^2(ay))^{4\beta_1}}\propto e^{-8a\beta_1 |y|} ,
\end{equation}
and for $s>1$
\begin{equation}
e^{2A_s(y)}=\frac{e^{-2\beta_s\tanh^{2s}\left(\frac{ay}{s}\right)}e^{\frac{s}{2(2s-1)}\beta_s {\sum_{n=1}^{s-1}\frac1{2n}\tanh^{2n}(\frac{ay}{s})}}}
{(\cosh^2(ay))^{\frac{2s}{(2s-1)}\beta_s}}\propto e^{\frac{-4a\beta_s}{(2s-1)} |y|} .
\end{equation}
Note that this represents a localized metric warp factor. It's worthwhile to observe the effect of deformation parameter $s$ in the energy density of the thick brane. The energy density is given by 
\begin{equation}
e(y)=T_{00}(y)=g_{00}\mathcal{L}-2\frac{\partial{\mathcal{L}}}{\partial g^{00}}.
\end{equation}

\begin{figure}[ht]
\centerline{\includegraphics[scale=0.8]{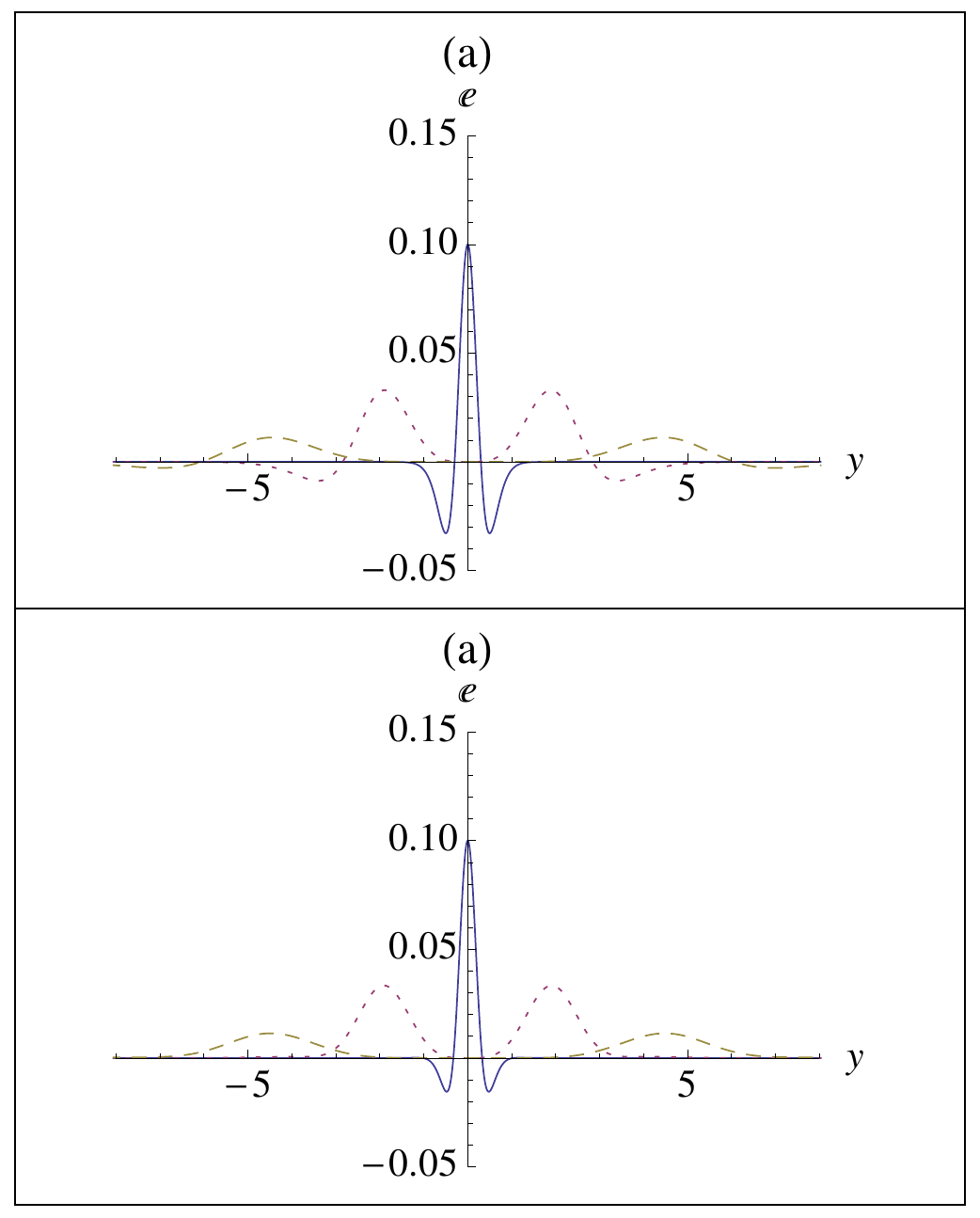}}
\caption{The energy density of the thick brane: (a) without dilaton and (b) with dilaton.  $s=1$ (lined) scaled by 1/10, $s=3$ (dotted), $s=5$ (dashed). }\label{energy-density}
\end{figure}

We show in Fig.\ref{energy-density} the energy density of the thick brane for various value of the deformation parameter $s$.

\section{Review of the Method}
In this section we give a review of the general prescription to be followed in all the cases to be studied in this work, for more details see\cite{Landim:2011ki}. 
First we take the equation of motion for the field in five dimensions, which generally takes the form
\begin{equation}
\hat{O}\Phi(x,y)=0, 
\end{equation} 
where $\hat{O}$ is a differential operator in five dimensions. The next step is to separate the differential operator in the brane ($\hat{O}_{4d}$) and extra dimension($\hat{O}_y$) and perform a separation of variables in the field $\Phi(x,y)=\psi(y)\phi(x)$. After some manipulations we arrive at two equations
$$
\hat{O}_{4d}\phi(x)=-m^2\phi(x)
$$
and
$$
\hat{O}_y\psi(y)=m^2\psi(y).
$$
The solutions of the second equation above give us the allowed masses in the visible brane. This equation in all cases will have the form
\begin{equation}
 \left(-\frac{d^2}{dy^2}+P'(y)\frac{d}{dy}+V(y)\right)\psi(y)=m^2Q(y)\psi(y)\label{pq}
\end{equation}
that can be transformed in a Schr\"odinger-like equation
\begin{equation}
 \left(-\frac{d^2}{dz^2}+\overline{U}(z)\right)\overline{\psi}(z)=m^2\overline{\psi}(z),\label{schlike}
\end{equation}
through the transformations
\begin{equation}
 \frac{dz}{dy}=f(y), \quad \psi(y)=\Omega(y)\overline{\psi}(z),
\end{equation}
with
\begin{equation}
 f(y)=\sqrt{Q(y)}, \quad \Omega(y)=\exp(P(y)/2)Q(y)^{-1/4},
\end{equation}
and
\begin{equation}
 \overline{U}(z)=V(y)/f^2+\left(P'(y)\Omega'(y)-\Omega''(y)\right)/\Omega f^2 \label{potential}.
\end{equation}
where the prime is a derivative with respect to $y$. 

The effective action is obtained by integrating in the extra dimension.  The finiteness of this action is the condition that $\psi$ is square integrable, just like in quantum mechanics with a potential Eq. (\ref{potential}). The study of the localization of massless modes ($m^2=0$) is very simple since is this case $\psi=\psi_0$ is a solution. In this way, we just need to analyze if the integral of the measure is finite. The massive mode case is more intricate because we generally have $\lim _{z \rightarrow \pm\infty}U(z)=0$ and we do not have a discrete spectrum for $m>0$. We must therefore consider plane waves coming from infinity and use the Transfer Matrix method to analyze the existence or possible resonances.

We will see that all potential of Eq. (\ref{schlike}) has the general form of Fig. \ref{fig:multistep2}
\FIGURE{
\centerline{\psfig{figure=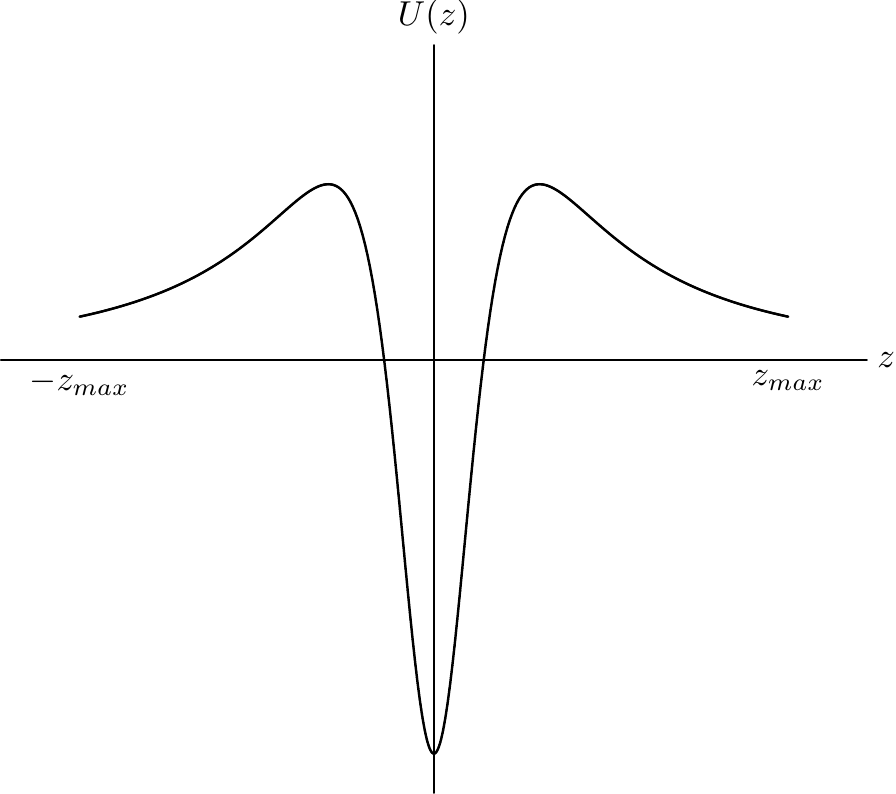,angle=0,height=5cm}}
\caption{General potential with parity symmetry with $\lim_{z \rightarrow \pm\infty}U(z)=0$. }\label{fig:multistep2}}
that can be approximated by a series of barriers. In each region showed in Fig. \ref{fig:multistep4} the Schr\"odinger equation can be solved for each interval $z_{i-1}<z<z_i$, where we have approximated the potential by 
\bq
U(z)=U(\overline{z}_{i-1})=U_{i-1},\quad\overline{z}_{i-1}=(z_i+z_{i-1})/2.
\ee
\FIGURE{
\centerline{\psfig{figure=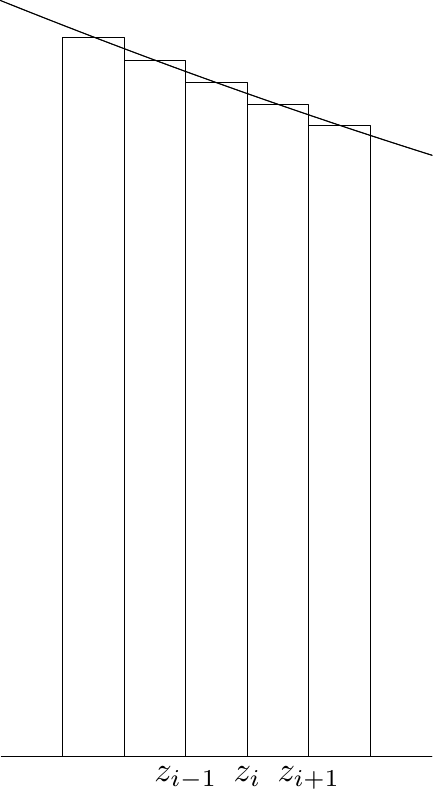,angle=0,height=6cm}}
\caption{The multistep regions.}\label{fig:multistep4}
}

The solution in this interval is
\bq
\psi_{i-1}(z)=A_{i-1}e^{ik_{i-1}z}+B_{i-1}e^{-ik_{i-1}z},\quad k_{i-i}=\sqrt{\lambda-U_{i-1}},
\ee
and the continuity of the $\psi_{i-1}(z)$ and $\psi'_{i-1}(z)$ at $z=z_i$ gives us
\bq
\left(
\begin{array}{c}
 A_i\\ 
B_i
\end{array} \right)
=
M_i\left(
\begin{array}{c}
 A_{i-1}\\ 
B_{i-1}
\end{array} \right).
\ee
In the above equation we have that
\bq
M_i=
\frac{1}{2k_i}\left[\begin{array}{cc}
 (k_i+k_{i-1})e^{-i(k_i-k_{i-1})z_i}& (k_i-k_{i-1})e^{-i(k_i+k_{i-1})z_i} \\ 
(k_i-k_{i-1})e^{i(k_i+k_{i-1})z_i} & (k_i+k_{i-1})e^{i(k_i-k_{i-1})z_i}
\end{array} \right]
\ee
and performing this procedure iteratively we reach
\bq
\left[
\begin{array}{c}
 A_N\\ 
B_N
\end{array} \right]
=
M\left(
\begin{array}{c}
 A_{0}\\ 
B_{0}
\end{array} \right),
\ee
where,
\bq
M=M_NM_{N-1}\cdots M_{2}M_1,
\ee
and the transmission coefficient is given by
\bq
T=1/|M_{22}|^2.
\ee
This expression is what we must to compute as a function of $\lambda$, which in our case is $m^2$. In order to obtain the numerical resonance values we choose the $z_{max}$ to satisfy $U(z_{max})\sim 10^{-4}$ and $m^2$ runs from $U_{min}=U(z_{max})$ to $U_{max}$ (the maximum potential value). We divide $2z_{max}$ by $10^4$ or $10^5$ such that we have $10^4+1$ or $10^5+1$ transfer matrices.

\section{The Deformed Gravity Resonances}

As said in the introduction we are not going to analyses the localization of zero modes of fields. The main objective is to look for the existence of possible resonances. This could indicate that this modes can be found at the visible brane. For completeness we consider the two backgrounds found in the last section.  In order to compute resonant massive modes we just need the Schr\"odinger-like equation that is obtained for each kind of field. Here we give a quick review of how to obtain this equation for the gravity field. More details can be found in Refs. \cite{Landim:2011ki,Kehagias:2000au}. 

We must consider the fluctuation $g_{MN}'=g_{MN}+h_{MN}$, where $h_{MN}$ represents the graviton in the axial gauge $h_{5M}=0$ and assume that
$h_{\mu}^{\mu}=\partial_{\mu}h^{\mu\nu}=0$. With these considerations we obtain the equation,
\begin{equation}\label{gravdilaton}
\left[-e^{2(A_s-B_s)}\frac{\partial^2}{\partial y^2}+e^{2(A_s-B_s)}B_s' \frac{d}{dy}+2e^{2(A_s-B_s)}\left(A_s^{''}-A_s'B_s'+2(A_s')^2\right)-\partial^2\right]h_{\mu\nu}=0,
\end{equation}
with the definition $\partial^2\equiv \eta^{\mu\nu}\partial_{\mu}\partial_{\nu}$. Performing now the separation of variables 
$h_{\mu\nu}= \overline{h}_{\mu\nu}(x)\psi(y)$, using the relation between $A_s$ and $B_s$, and our previous transformations Eq. (\ref{potential}) we get the Schr\"odinger equation
\begin{equation}
\left(-\frac{d}{dz^2}+\overline{U}\right)\overline{\psi} =m^2\overline{\psi},
\end{equation}
with potential given by
\begin{equation}
\overline{U}=\frac{3}{2}e^{3A_s/2}\left(A_s^{''}+\frac{9}{4}(A_s')^2\right).
\end{equation}

Now we consider the case with a dilaton free background. The strategy here is the same as that used to find the background solution without the dilaton contribution. The final result is obtained just by setting $B_s=0$ in 
Eq. (\ref{gravdilaton}) and after following the same steps we arrive at the Schr\"odinger equation with potential given by
\begin{equation}
\overline{U}(z)=\frac{3}{4}e^{2A_s}\left(2A_s^{''}+5(A_s')^2\right).
\end{equation}

The case $s=1$ has been considered previously and we show in Fig. \ref{fig:gravs35} the graphics of the gravity potential for $s=3,5$. The behavior $\lim _{z \rightarrow \pm\infty}U(z)=0$ show that far of the brane we have
plane waves. We must compute transmission coefficients to analyze the possibility of the existence of some gravity massive mode in the brane. We use the method of transfer matrix proposed previously. 
We show in Figs. \ref{fig:grav-logts3} and \ref{fig:grav-logts5} the results of $\log T$.

\FIGURE{
\centerline{\psfig{figure=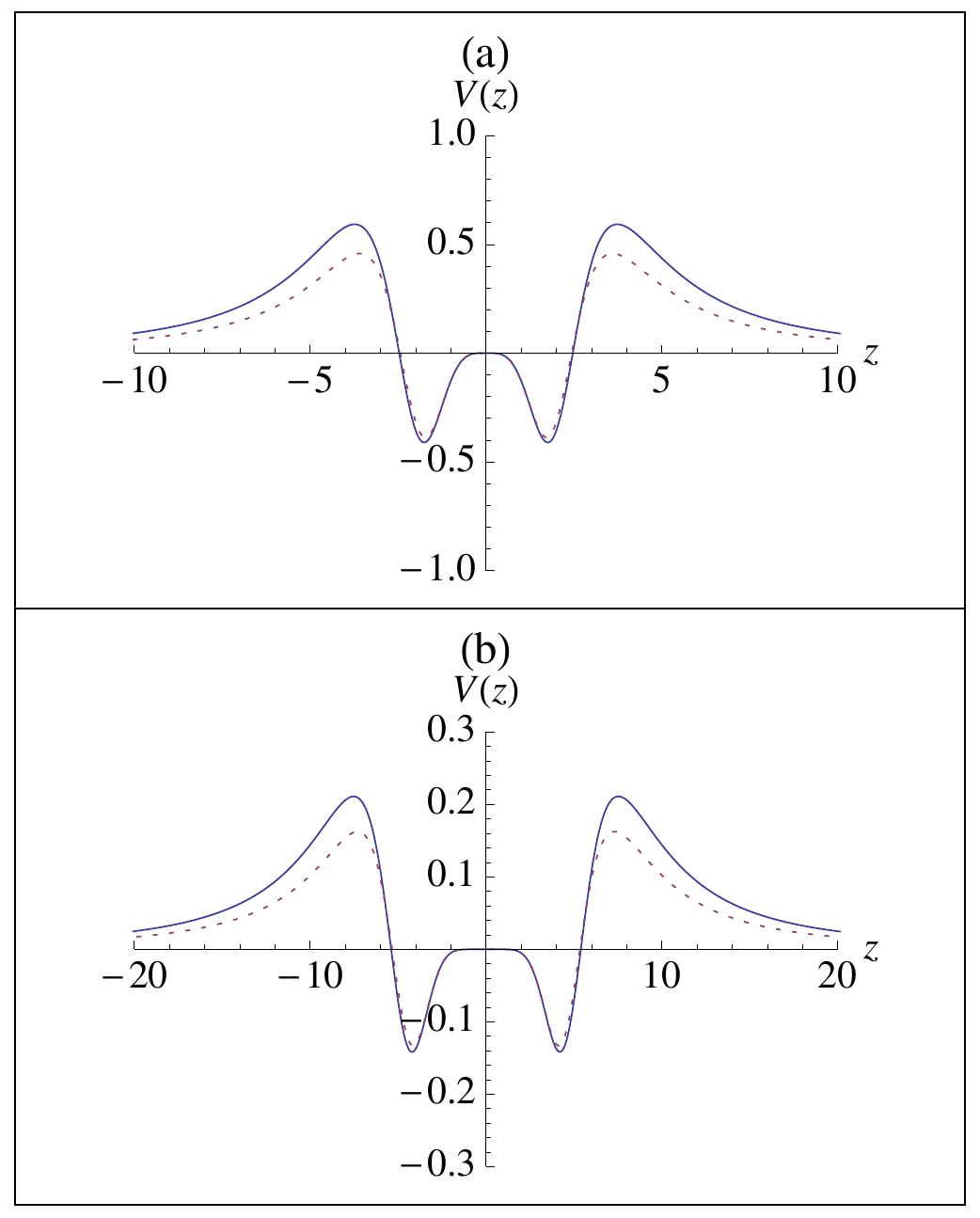,angle=0,height=7cm}}
\caption{Potential of the Schr\"odinger like equation of the gravitational field with dilaton (lined) and without dilaton (dotted) for $s=3$ (a) and $s=5$ (b).\label{fig:gravs35}}
}

\FIGURE{
\centerline{\psfig{figure=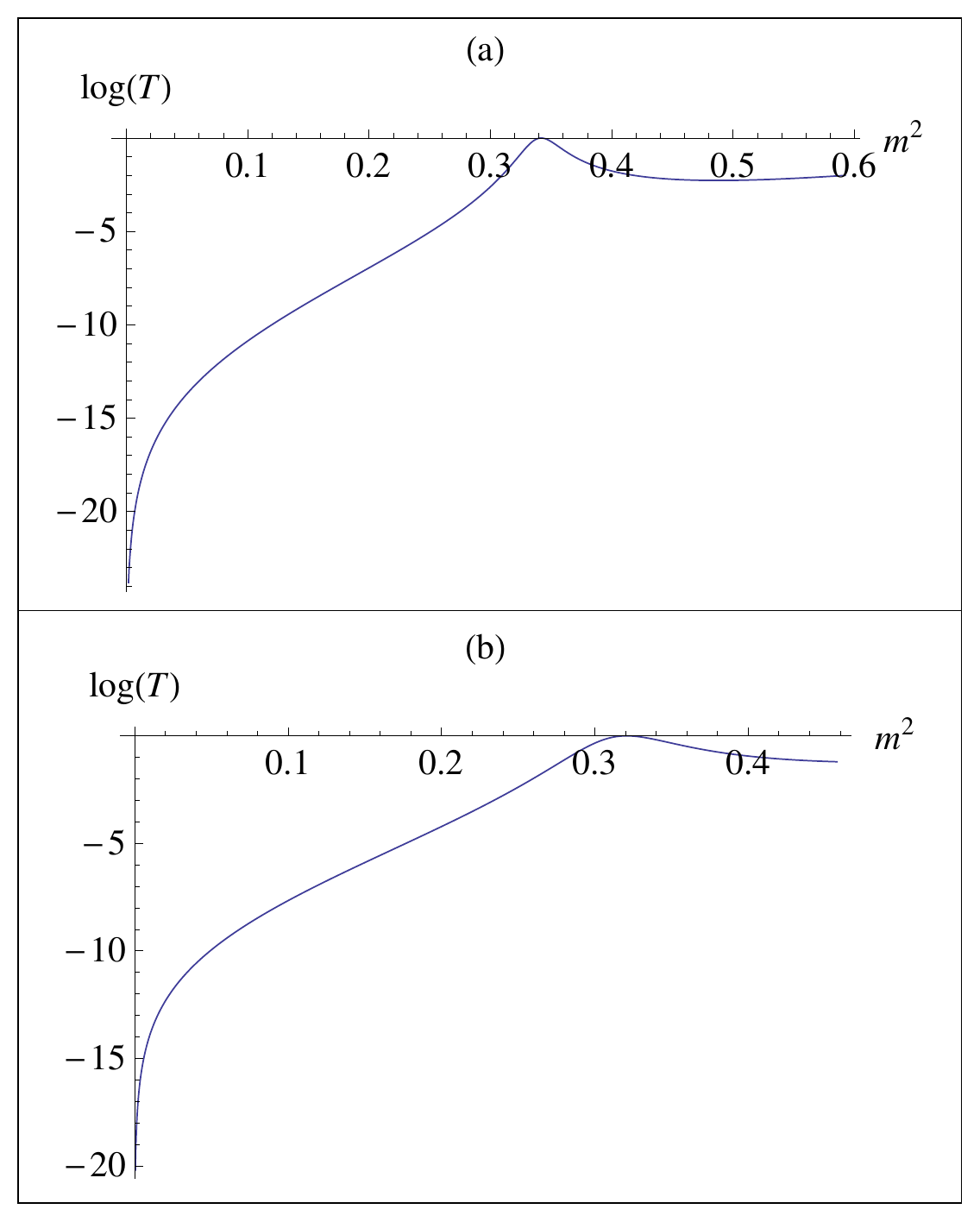,angle=0,height=7cm}}
\caption{Logarithm of the transmission coefficient of  gravitational field for $s=3$ (a) without dilaton  and (b)with dilaton.\label{fig:grav-logts3}}
}

\FIGURE{
\centerline{\psfig{figure=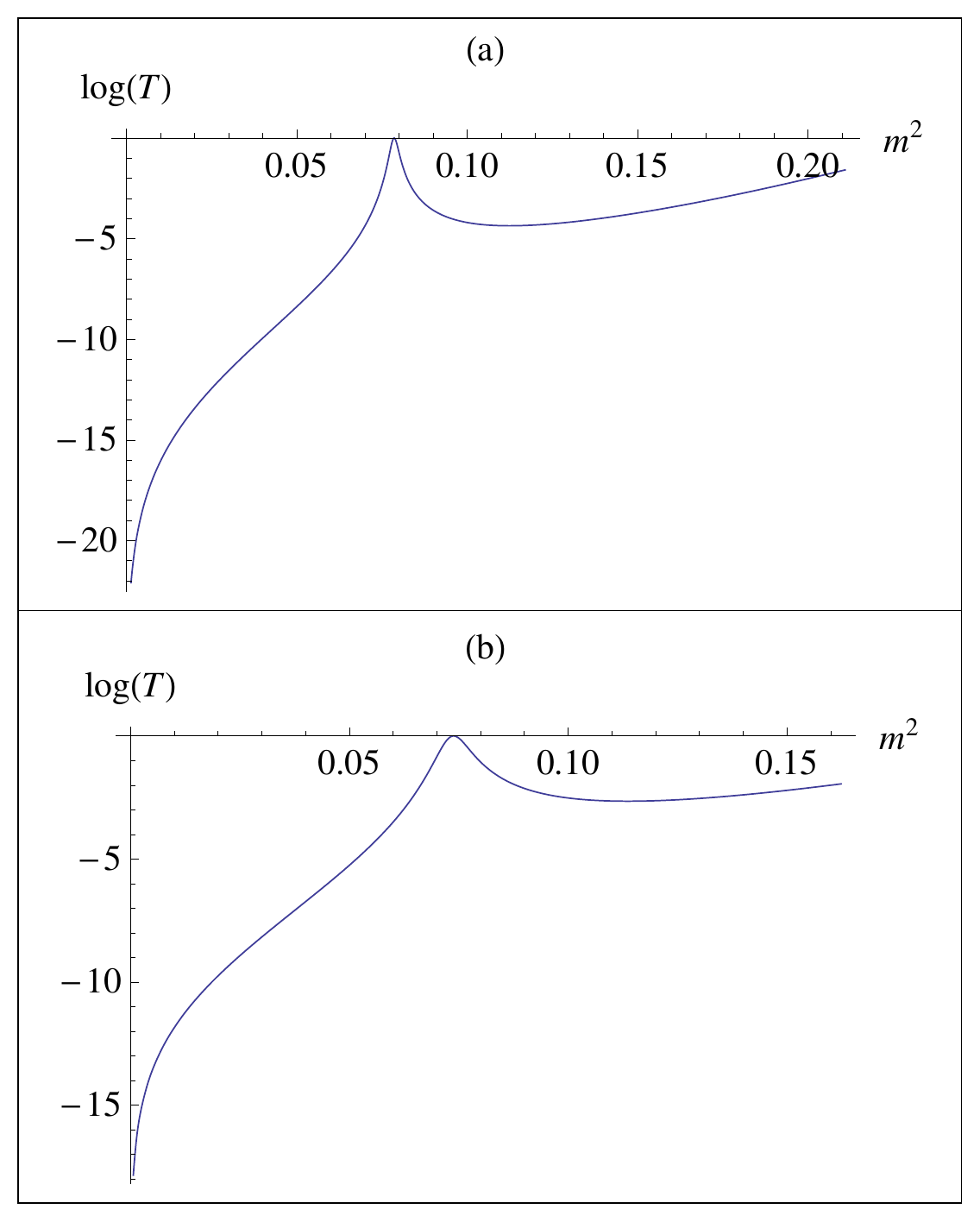,angle=0,height=7cm}}
\caption{Logarithm of the transmission coefficient of the gravitational field for $s=5$ (a) without dilaton  and (b)with dilaton.\label{fig:grav-logts5}}
}

The results obtained enforces the previous affirmation that the resonances are strongly dependent on the topological defect chosen. Differently of the case $s=1$, for $s=3$ even without the dilaton, we have
a peak of resonance at $m^2\approx 0,32$ as can be seen form Fig.\ref{fig:grav-logts3} (a). In Fig.\ref{fig:grav-logts3} (b) we see that the effect of the dilaton is to change the value of the resonance to $m^2\approx 0,31$. 
In Figs. \ref{fig:grav-logts5} (a) and (b) we see again that, for $s=5$ we have resonances regardless the presence of the dilaton. This is an interesting result since apparently the cases with $s\neq 1$ all present resonances 
for the gravity field.

\section{The Deformed Scalar Field Resonances}

In this section we study the massive modes of scalar fields. This can be considered the simplest application of the method. We follow the same steps as in the gravity field and review how to obtain the 
Schr\"odinger like equation. 

Initially we consider the action for the scalar field $\Phi$ coupled to
gravity,
\begin{equation}\label{acao}
\frac{1}{2}\int d^4xdy \sqrt{-g}e^{-\lambda\pi}g^{MN}\partial_M\Phi\partial_N\Phi,
\end{equation}
where the indexes $M,N$ go from $0$ to $4$. The equations of motion are
\begin{equation}\label{scalardilaton}
\eta^{\mu\nu}\partial_\mu\partial_\nu\Phi+e^{-2A_s-B_s+\lambda\pi}\partial_y[e^{4A_s-B_s-\lambda\pi}\partial_y\Phi]=0,
\end{equation}
and with the separation of variables
\begin{equation}\label{eq}
\Phi(x,y)=\chi(x)\psi(y),
\end{equation}
we arrive at the following equation for $\psi (y)$
\begin{equation}
\{\frac{d^2}{dy^2}+(4A_s'-B_s'-\lambda\pi)\frac{d}{dy}\}\psi=-m^2e^{2(B_s-A_s)}\psi.\label{ydep}
\end{equation}

Now using the relation between $B_s$ and $A_s$ and the transformation (\ref{potential}) we get the Schr\"odinger equation
\begin{equation}\label{schrop}
\left\{-\frac{d^2}{dz^2}+\overline{U}(z)\right\}\overline{\psi}=m^2\overline{\psi},
\end{equation}
where the potential $\overline{U}(z)$ assumes the form
\begin{eqnarray}
 &&\overline{U}(z)=e^{3A_s/2}\left((\frac{\alpha^2}{4}-\frac{9}{64})A_s'(y)^2-(\frac{\alpha}{2}+\frac{3}{8})A_s''(y)\right),  \nonumber 
\end{eqnarray}
where $\alpha=-15/4-\lambda\sqrt{3M^3}$.

For the dilaton free case we just fix $B_s=\pi=0$ in Eq. \ref{scalardilaton} and after performing the separation of variables and the transformation Eq. (\ref{potential}) 
we arrive at the Schr\"odinger equation with potential $\overline{U}(z)$ given by 
\begin{equation}
\overline{U}(z)=e^{2A_s}\left[\frac{15}{4}(A_s')^2+\frac{3}{2}A_s''\right].
\end{equation}

We show in Fig. \ref{fig:s35q0} the plots for the potential of the associated Schr\"odinger equation. Both looks very similar to a double barrier and we should expect the existence of resonant modes. In fact in 
Figs. \ref{fig:logts3q0} and \ref{fig:logts5q0} we see the graphic of the logarithm of the transmission coefficient for the deformation parameters $s=3,5$, with and without the dilaton respectively. It is important to note that, differently from the gravity field, the presence of the dilaton induces a richer structure of resonances. We can see in Figs. \ref{fig:logts3q0} (a) and \ref{fig:logts5q0} (a) that the presence of the dilaton 
gives us $15$ and $20$ peaks of resonance for $s=3$ and $5$ respectively. This indicates that, for this model, many massive scalar field have a peak of probability to interact with the membrane.

\FIGURE{
\centerline{\psfig{figure=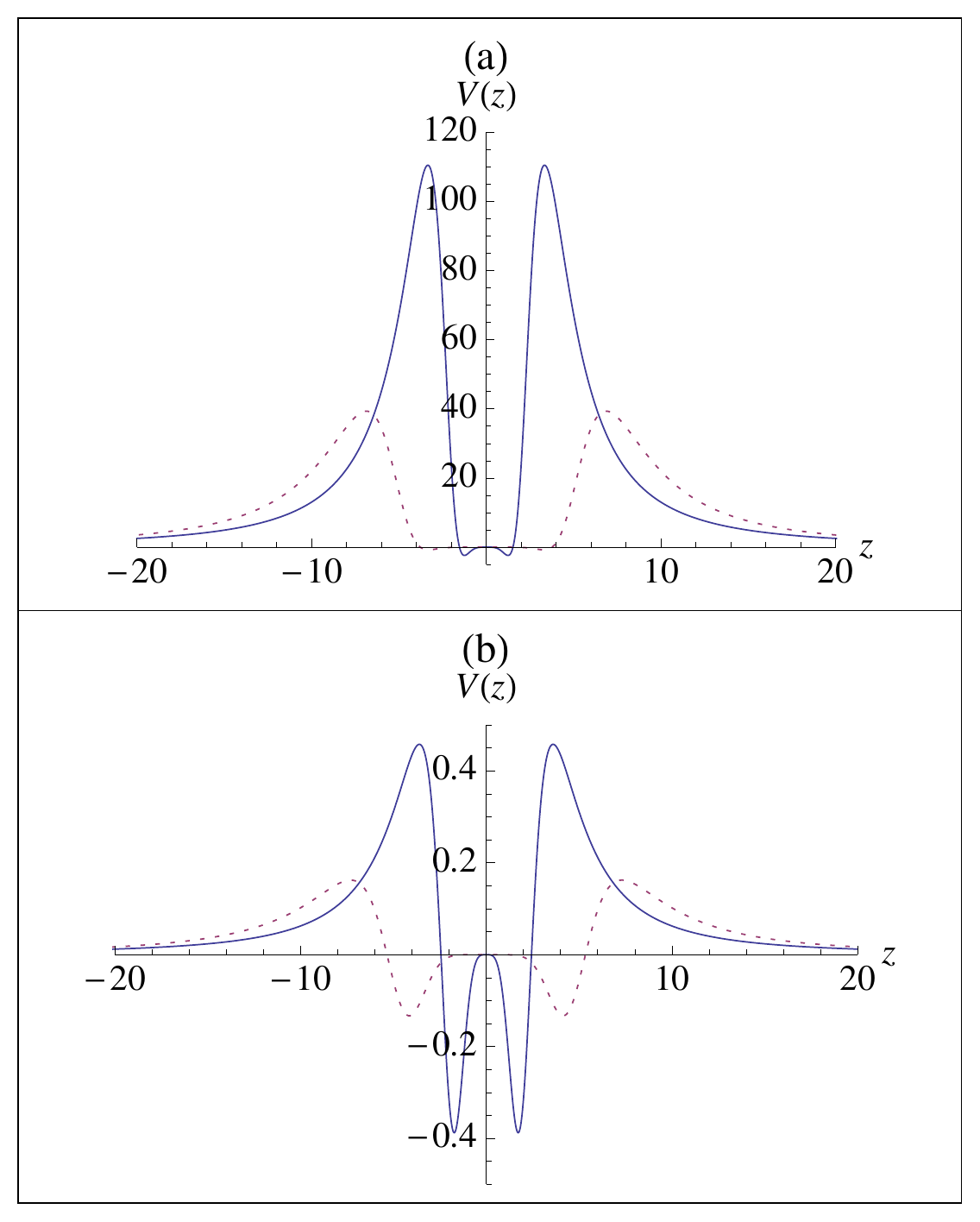,angle=0,height=8cm}}
\caption{Potential of the Schr\"odinger like equation of scalar field for $s=3$ (lined) and $s=5$ (dotted). (a) with dilaton for $\lambda\sqrt{3M^3}=40$ and (b) without dilaton.\label{fig:s35q0}}
}

\FIGURE{
\centerline{\psfig{figure=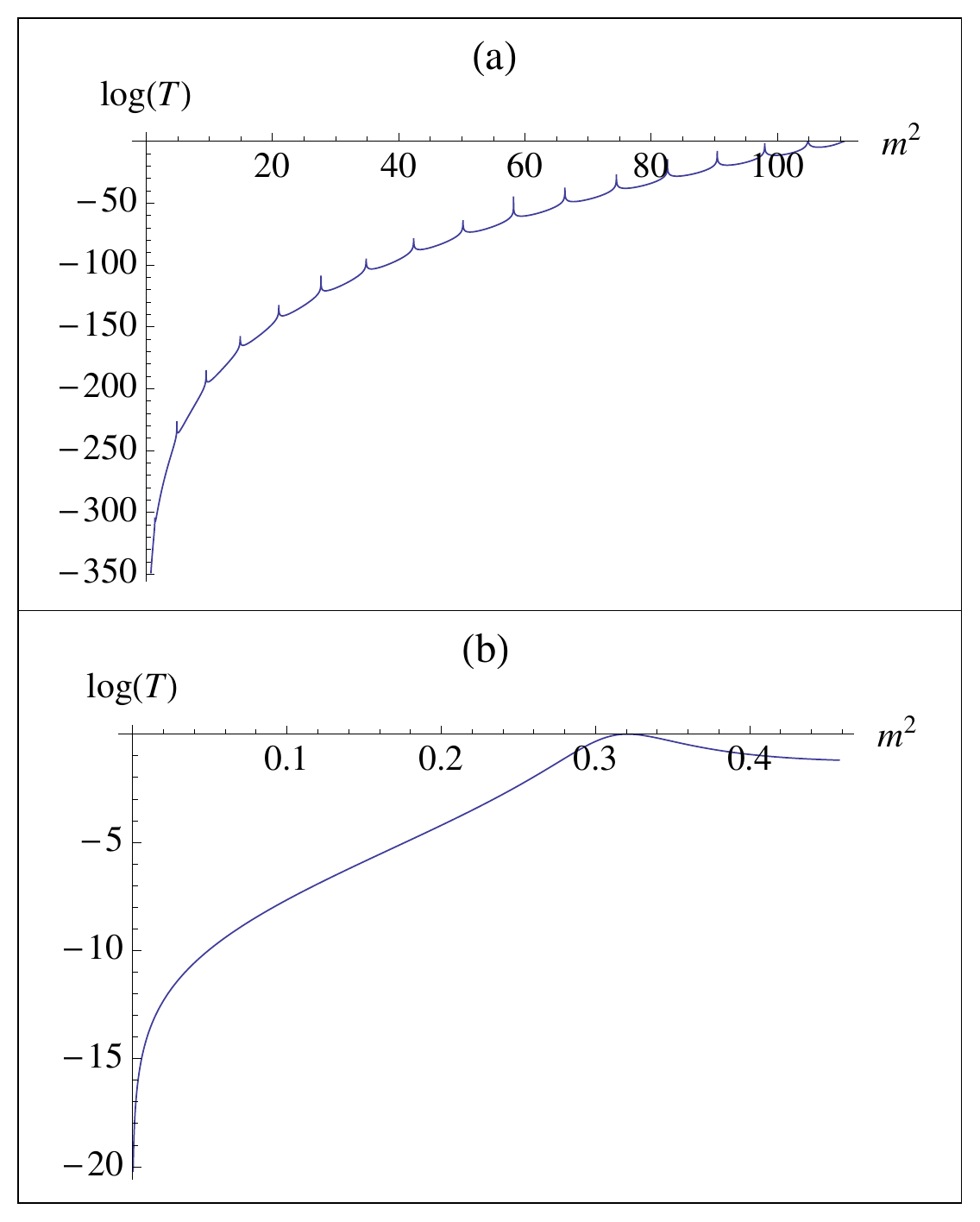,angle=0,height=8cm}}
\caption{Logarithm of the transmission coefficient of the scalar field for $s=3$. (a) with the dilaton for $\lambda\sqrt{3M^3}=40$, (b) without the dilaton.\label{fig:logts3q0}}
}

\FIGURE{
\centerline{\psfig{figure=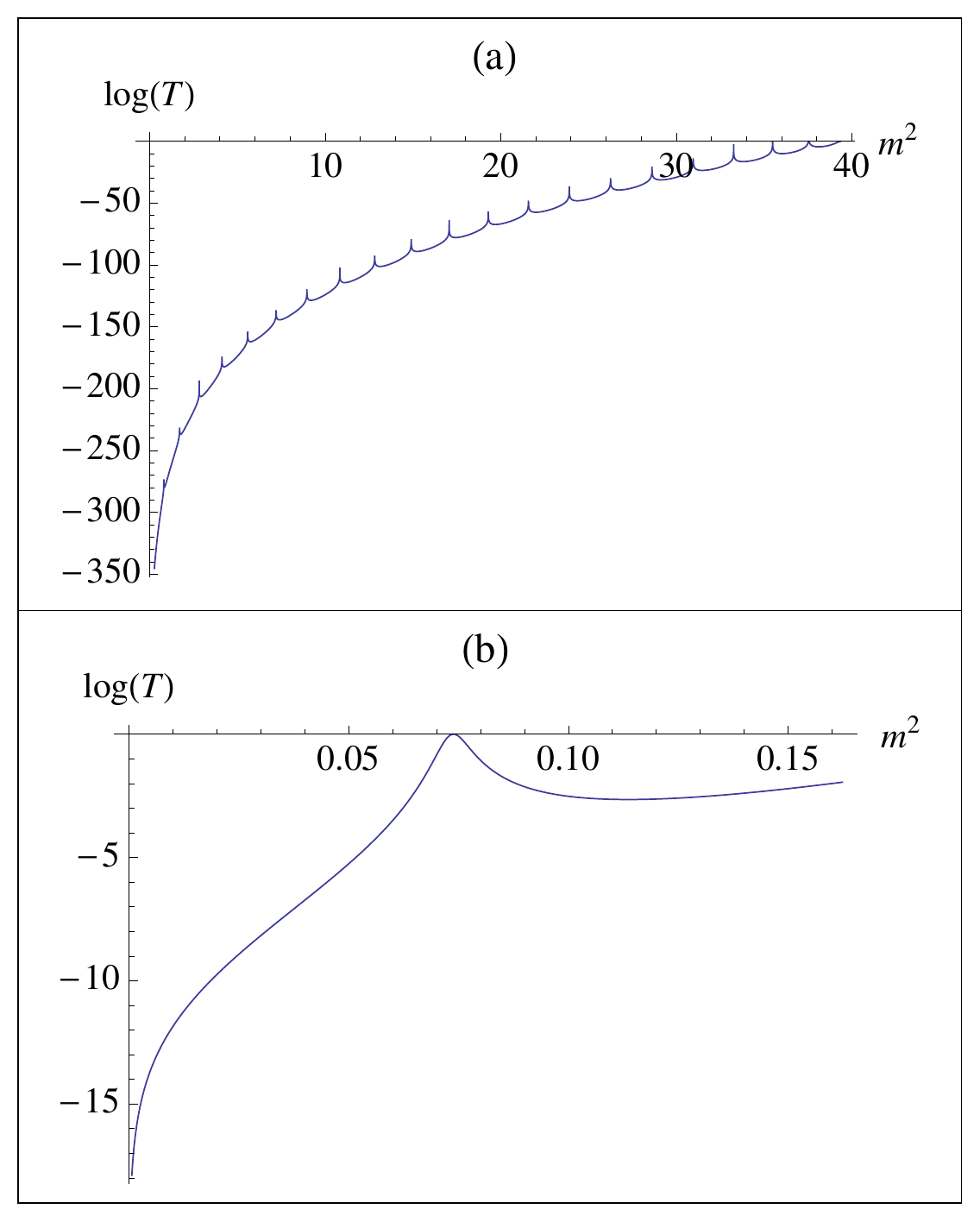,angle=0,height=8cm}}
\caption{Logarithm of the transmission coefficient of the scalar field for $s=5$. (a) with the dilaton for $\lambda\sqrt{3M^3}=40$, (b) without the dilaton.\label{fig:logts5q0}}
}

\section{The Deformed Vector Field Resonances}

Here we consider the next important field, which is the gauge field. The undeformed case was studied previously and now we consider the richer deformed case. For completeness, we give an analyzes of the resonances  for both cases: with and without the dilaton. They give different phenomenological results, that is why is important to consider the two cases. These cases have already been studied in \cite{Kehagias:2000au}, but we analyze it here in the light of the Transfer Matrix Method. We do not give the full details of how to arrive at the potential of the Schr\"odinger-like equation since this can be found in Ref. \cite{Landim:2011ki}. 

Consider the action of the gauge field coupled wit the dilaton
\begin{equation}
S_X=\int d^{5}x\sqrt{-g}e^{-\lambda \pi }[Y_{M_{1}M_{2}}Y^{M_{1}M_{2}}],
\end{equation}
where $Y_{M_{1}M_{2}}=\partial _{[M_{1}}X_{M_{2}]}$ is the field
strength for the $1$-form $X$. The equation of motion reads
\begin{equation}
\partial_{M}(\sqrt{-g}g^{MP}g^{NQ}e^{-\lambda\pi}H_{PQ})=0.
\end{equation}

We use the gauge freedom to fix $X_{y}=\partial^{\mu}X_{\mu}=0$ to simplify this equation and arrive at
\begin{eqnarray}\label{umformadilaton}
\partial _{\mu_1 }Y^{\mu_1\mu_{2}}
+e^{((-B_s+\lambda\pi)}\partial _{y}[e^{(2A_s-B_s-\lambda\pi)}{X'}^{\mu_2}]=0,
\end{eqnarray}
and performing the separation of variable we get the equation for the $\psi (y)$ dependence
\begin{equation}
\{\frac{d^2}{dy^2}-\left( -2{A_s}'+{B_s}'+\lambda\pi'\right)\frac{d}{dy}\}\psi\left(
y\right) =-m^{2}e^{2\left( B_s-A_s\right) }\psi\left( y\right) . \label{Udilaton}
\end{equation}
Now we can find the Schr\"odinger equation via the transformation Eq. (\ref{potential}) and use the relations between $B_s,\pi$ and $A_s$. We arrive at
\begin{eqnarray}
 &&\overline{U}(z)=e^{3A_s/2}\left((\frac{\alpha^2}{4}-\frac{9}{64})A_s'(y)^2-(\frac{\alpha}{2}+\frac{3}{8})A_s''(y)\right),  \nonumber ,
\end{eqnarray}
where $\alpha=-7/4-\lambda\sqrt{3M^3}$.

Now we repeat the same strategy used throughout all the paper to reach the dilaton free case. We fix $B_s=\pi=0$ in Eq. \ref{umformadilaton} and perform the separation of variables and use the transformation
(\ref{potential}) to get the potential
\begin{equation}
\overline{U}=e^{2A_s}\{\frac{3}{4}A_s'^2+\frac{1}{2}A_s''\}.
\end{equation}
of the Schr\"odinger like equation.

We show in the Fig.\ref{fig:s35q1} plots of the vector field potential. We can see from this that in the presence of the dilaton the potential looks similar to a double barrier and we should have more 
resonances than in the case without the dilaton. In fact this can be seem in Figs.\ref{fig:logts3q1} and \ref{fig:logts5q1}, where  we show the logarithm of the transmission coefficient for $s=3$ and $5$ respectively. 
From Figs.\ref{fig:logts3q1} (b) and \ref{fig:logts5q1} (b) we do not have resonances. We should point that in the case $s=1$ without the dilaton no resonances has been found \cite{Landim:2011ki}. Therefore, 
the increase of $s$ induces more resonant modes, as mentioned in the introduction. With the dilaton we confirmed our expectation that the presence of the dilaton also increases the number of resonances.  
For this case we have found previously for the case $s=1$ six resonances \cite{Landim:2011ki}. Here, we can see in Figs. Figs.\ref{fig:logts3q1} (a) and \ref{fig:logts5q1} (a) that for $s=3$ and $5$ we have $15$ and $19$ 
peaks respectively.  
\FIGURE{
\centerline{\psfig{figure=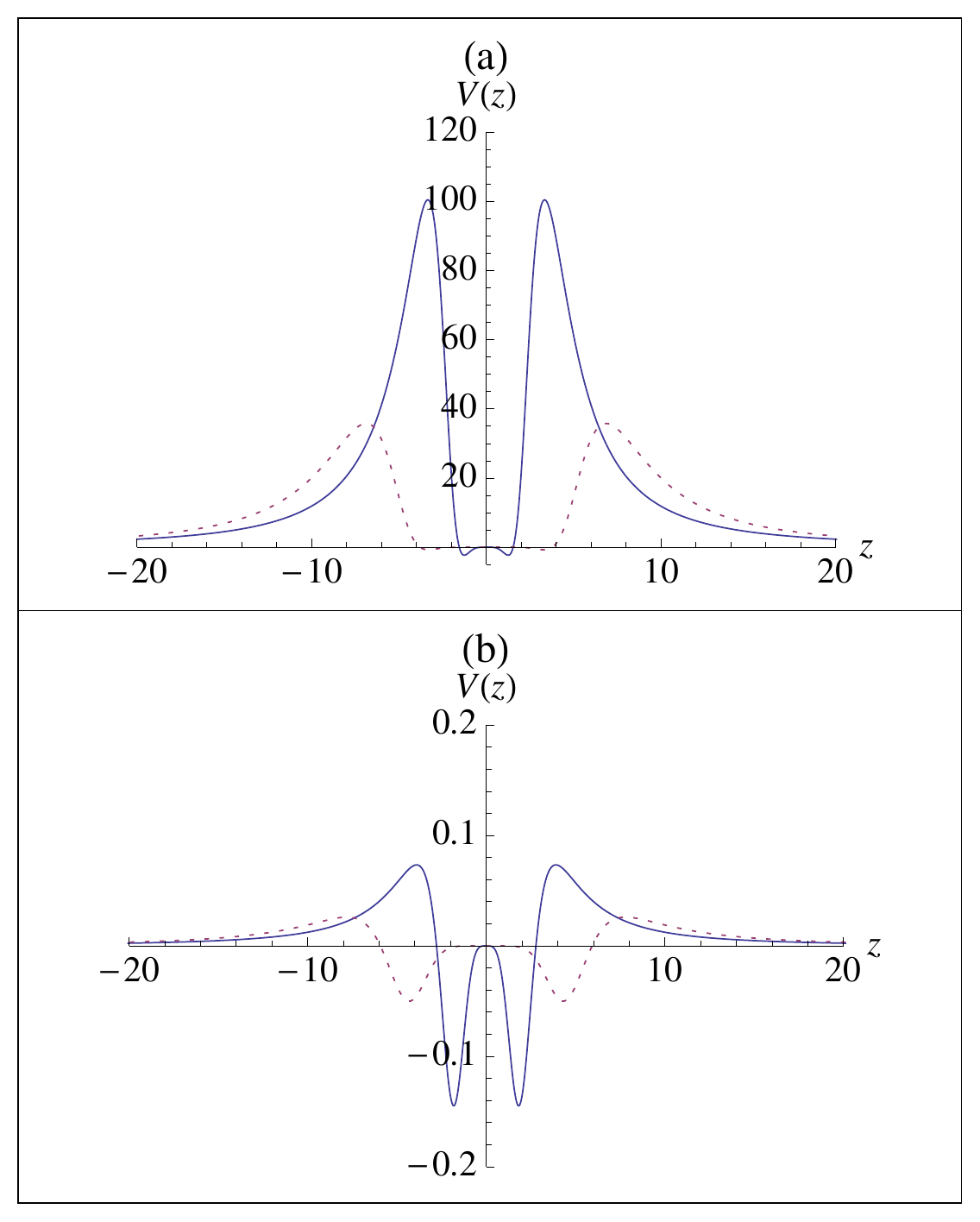,angle=0,height=8cm}}
\caption{Potential of the Schr\"odinger like equation of the vector field for $s=3$ (lined) and $s=5$ (dotted). (a) with dilaton for $\lambda\sqrt{3M^3}=40$ and (b) without dilaton.\label{fig:s35q1}}
}

\FIGURE{
\centerline{\psfig{figure=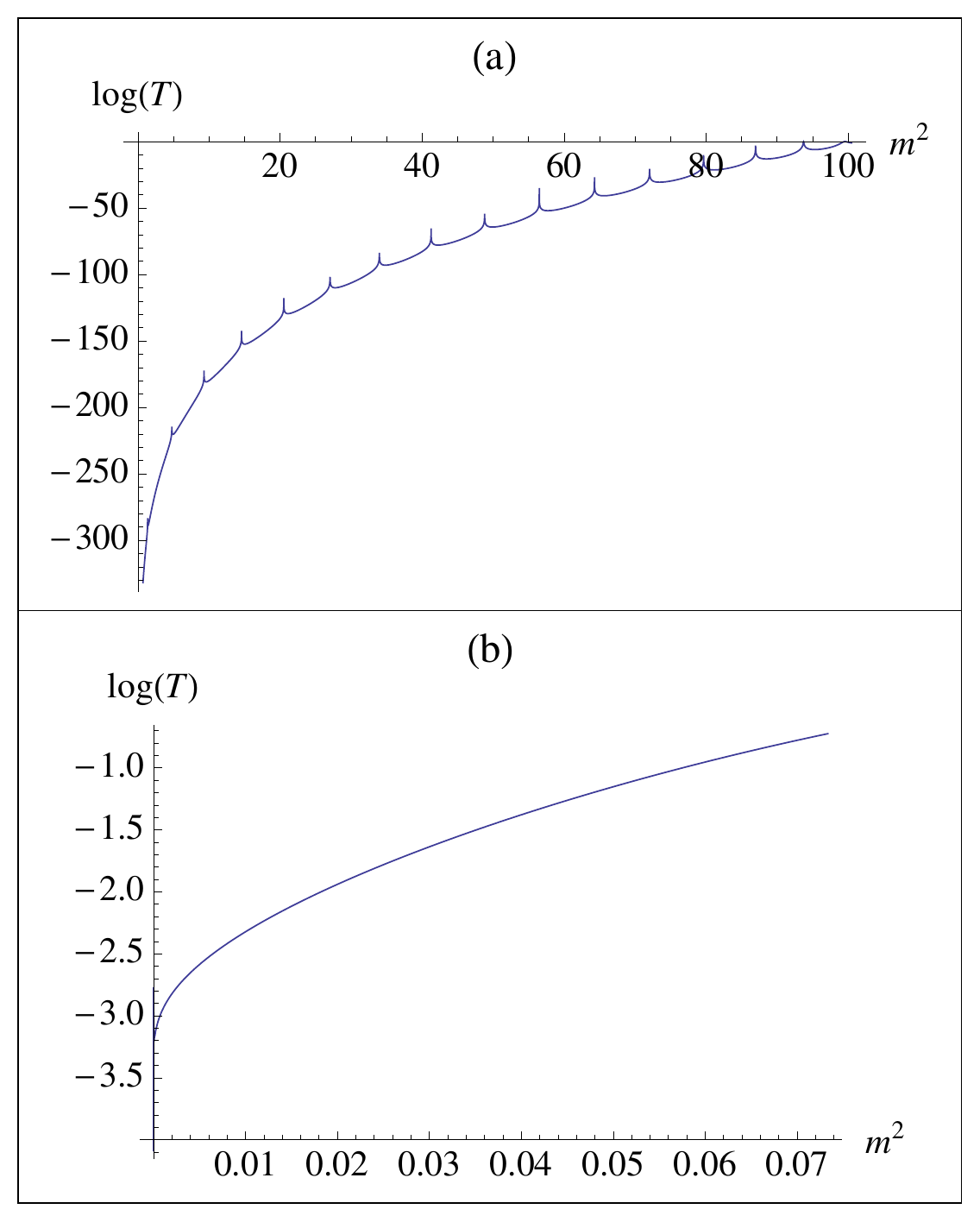,angle=0,height=8cm}}
\caption{Logarithm of the transmission coefficient of the vector field for $s=3$. (a) with the dilaton for $\lambda\sqrt{3M^3}=40$, (b) without the dilaton.\label{fig:logts3q1}}
}

\FIGURE{
\centerline{\psfig{figure=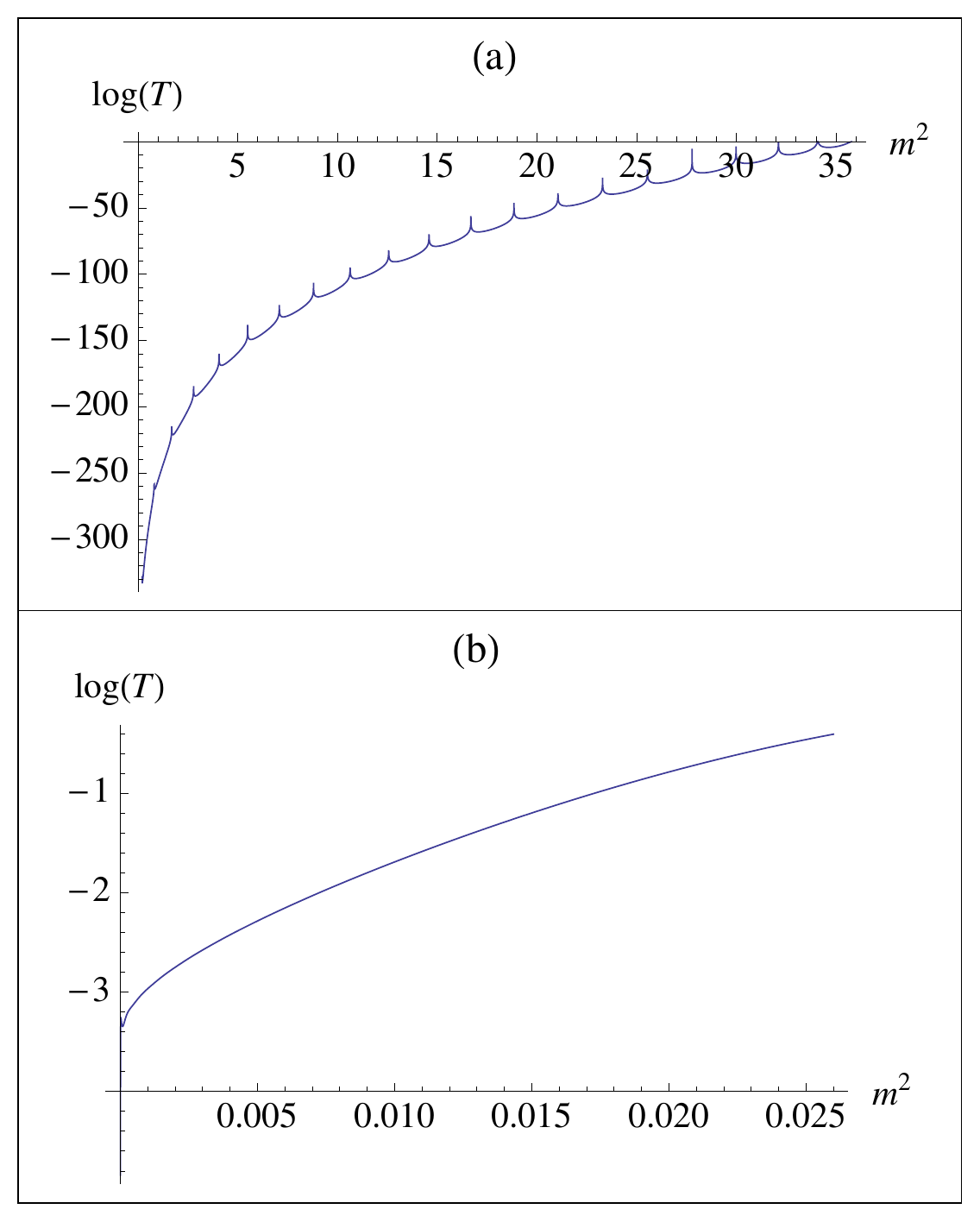,angle=0,height=8cm}}
\caption{Logarithm of the transmission coefficient of the vector field for $s=5$. (a) with the dilaton for $\lambda\sqrt{3M^3}=40$, (b) without the dilaton.\label{fig:logts5q1}}
}

\section{The Deformed Kalb-Ramond Resonances}

The dilaton can be seen as the radius of a compactified extra dimension and all the fields get an exponential factor coming from the measure. Hence, the same coupling for the scalar and vector fields can be used. More details about how to obtain these equations can be found in \cite{Landim:2011ki} and we give here just a short review.  Consider the action of the coupled system
\begin{equation}
S_X=\int d^{5}x\sqrt{-g}e^{-\lambda \pi }Y_{M_{1}M_{2}M_{3}}Y^{M_{1}M_{2}M_{3}},
\end{equation}
where $Y_{M_{1}M_{2}M_{3}}=\partial _{[M_{1}}X_{M_{2}M_{3}]}$ is the field strength for the $2$-form $X$. The equation of motion is given by
\begin{equation}
\partial_{M}(\sqrt{-g}g^{MP}g^{NQ}g^{LR}e^{-\lambda\pi}H_{PQR})=0.
\end{equation}

Gauge freedom can be used to fix $X_{\mu_{1}y}=\partial^{\mu_{1} }X_{\mu{_{1}}\mu_{2}}=0$ and the equation is simplified to
\begin{eqnarray}\label{doisformadilaton}
\partial _{\mu_{1} }Y^{\mu_{1} \mu_{2}\mu_{3}}
+e^{(2A_s-B_s+\lambda\pi)}\partial _{y}[e^{-(B_s+\lambda\pi)}{X'}^{\mu_{2}\mu_{3}}]=0,
\end{eqnarray}
and we can now separate the $y$ dependence of the field to get the equation for $\psi (y)$
\begin{equation}\label{zero}
\frac{d^{2}\psi(y)}{dy^{2}}-(\lambda\pi^{\prime}(y)+B_s^{\prime}(y))\frac{d\psi(y)}{dy}=-m^{2}e^{2(B_s(y)-A_s(y))}\psi(y).
\end{equation}
Using the previous relation between $B_s$,$\pi$ and $A_s$ and performing the transformation (\ref{potential}) we get
\begin{equation}\label{schro}
\left\{-\frac{d^2}{dz^2}+\overline{U}(z)\right\}\overline{\psi}=m^2\overline{\psi},
\end{equation}
with potential $\overline{U}(z)$ given by
\begin{equation}\label{pot_reson}
\overline{U}(z)=e^{\frac{3}{2}A_s}\left[\left(\frac{\alpha^2}{4}-\frac{9}{64}\right)(A_s')^2-\left(\frac{\alpha}{2}+\frac{3}{8}\right)A_s''\right].
\end{equation}
where,
\begin{equation}
\beta=\frac{\alpha}{2}+\frac{3}{8},\,\alpha=\frac{1}{4}-\sqrt{3M^3}\lambda.
\end{equation}

For the case without the dilaton we can again obtain the equations of motion setting $B_s=\pi=\lambda=0$ in \ref{doisformadilaton}. Following the same
reasoning as in all cases before we reach the potential for the Schr\"odinger equation
\begin{equation}
\overline{U}(z)=e^{2A_s}[\frac{1}{4}(A_s')^2+\frac{1}{2}(A_s'')].
\end{equation}

We show in the Fig.\ref{fig:s35q2} the graphics of the Kalb-Ramond field potential with and without the dilaton. In Figs.\ref{fig:logts3q2} and \ref{fig:logts5q2} we show the logarithm of the transmission coefficient for $s=3$ and $5$ respectively. Just like in the scalar and vector field cases we see that the spectrum of resonances is richer if we add the dilaton or increase the value of $s$. From the figures is easy to see that the presence of the dilaton induces more resonances. For the case with the dilaton we also see that increasing the value of $s$ the number of peaks also increases. We can see that for $s=1$ \cite{Landim:2011ki} we had $5$ peaks and from
Figs.\ref{fig:logts3q2} (a) and \ref{fig:logts5q2} (a) we have $14$ and $18$ peaks for $s=3$ and $5$ respectively. 
 
\FIGURE{
\centerline{\psfig{figure=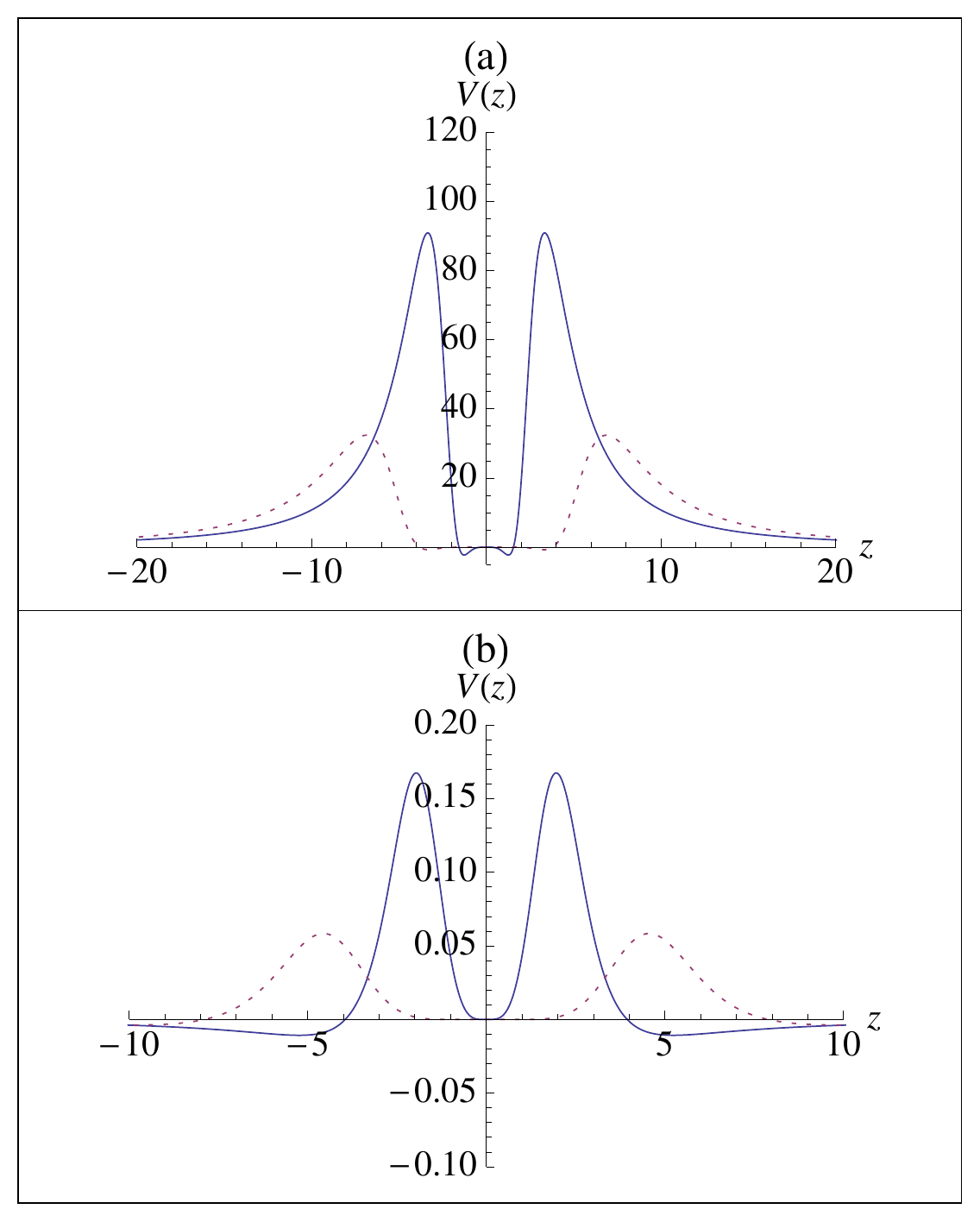,angle=0,height=8cm}}
\caption{Potential of the Schr\"odinger like equation of the Kalb-Ramond field for $s=3$ (lined) and $s=5$ (dotted). (a) with dilaton for $\lambda\sqrt{3M^3}=40$ and (b) without dilaton.\label{fig:s35q2}}
}

\FIGURE{
\centerline{\psfig{figure=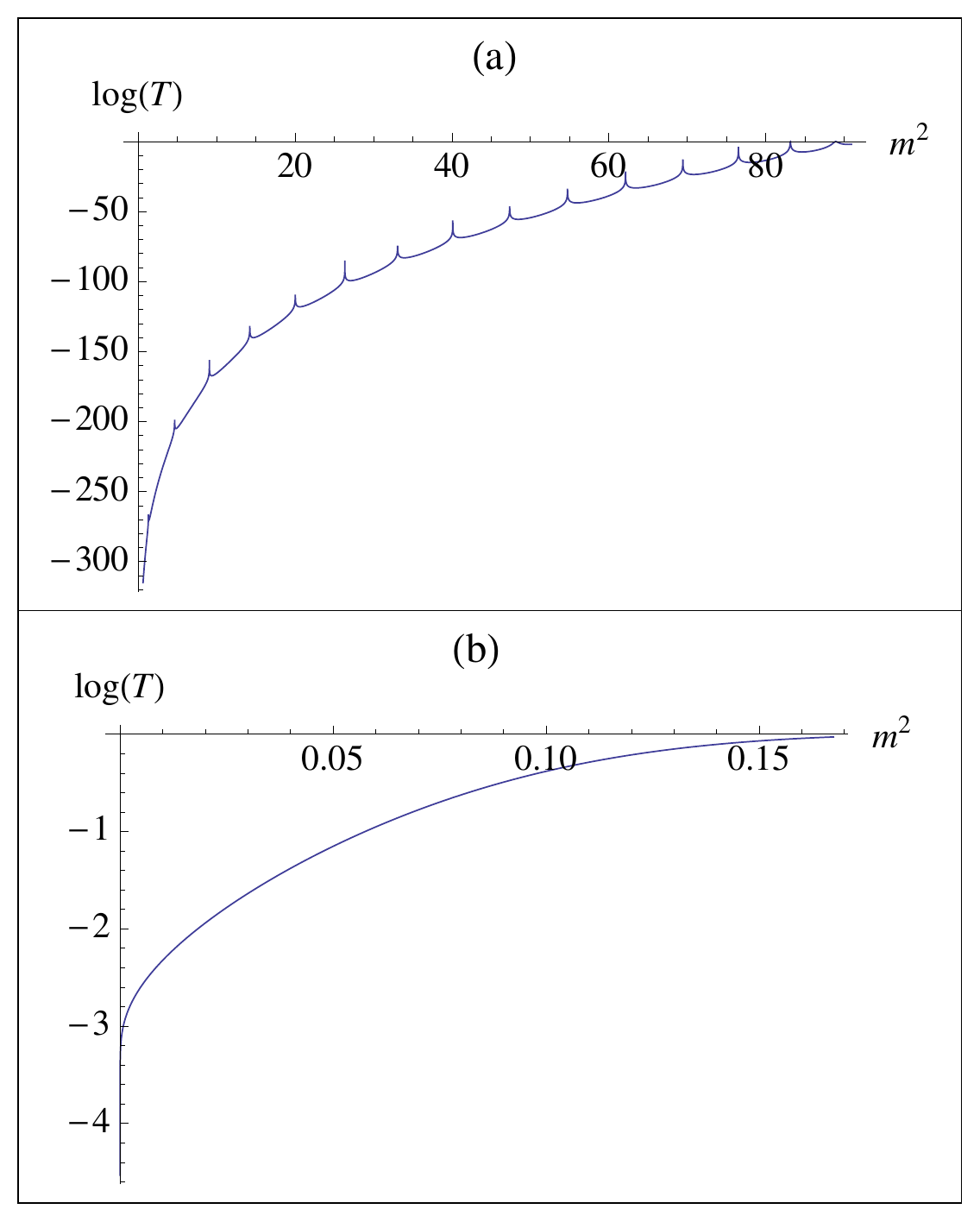,angle=0,height=8cm}}
\caption{Logarithm of the transmission coefficient of the Kalb-Ramond field for $s=3$. (a) with the dilaton for $\lambda\sqrt{3M^3}=40$, (b) without the dilaton.\label{fig:logts3q2}}
}

\FIGURE{
\centerline{\psfig{figure=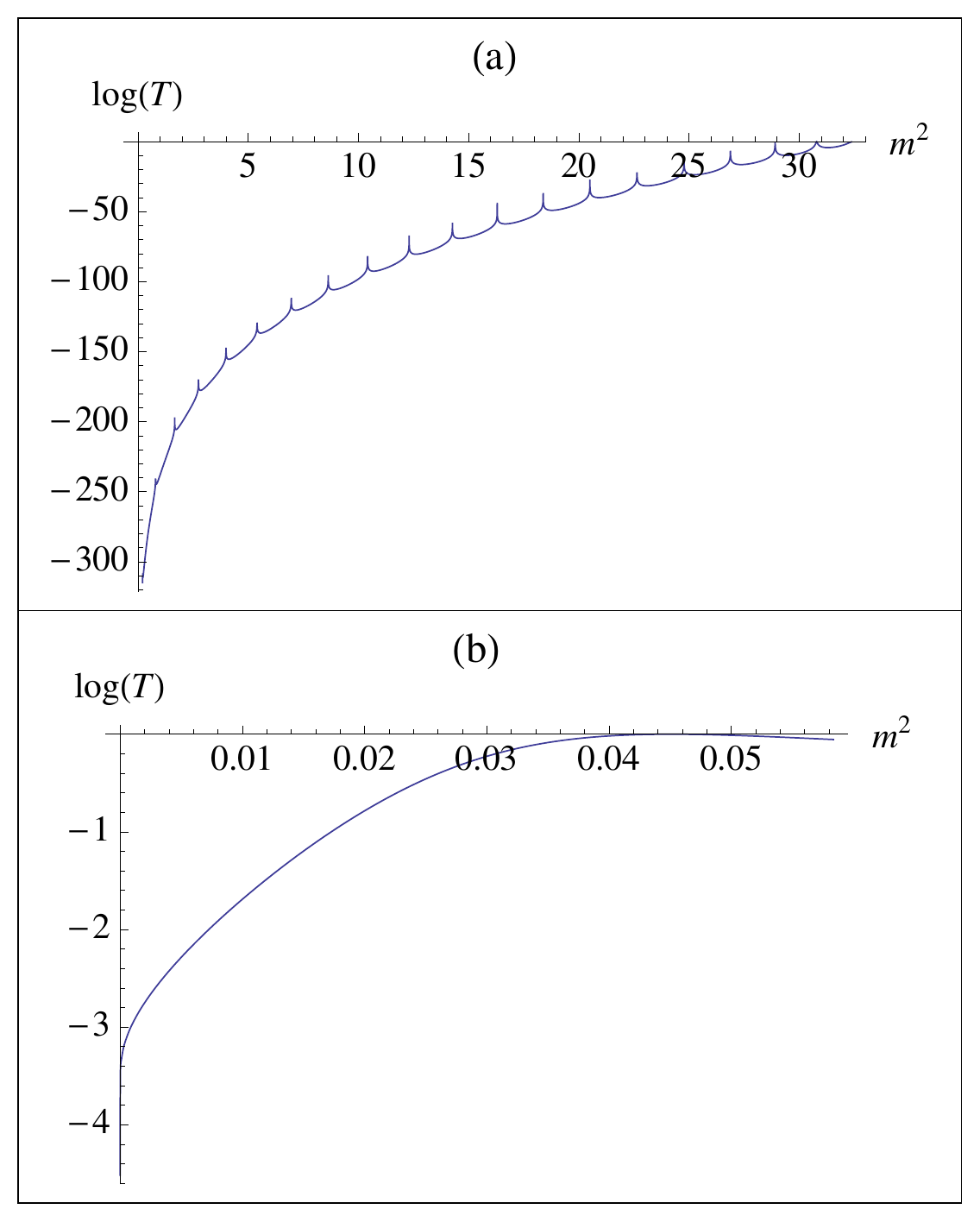,angle=0,height=8cm}}
\caption{Logarithm of the transmission coefficient of the Kalb-Ramond field for $s=5$. (a) with the dilaton for $\lambda\sqrt{3M^3}=40$, (b) without the dilaton.\label{fig:logts5q2}}
}

\section{The Deformed Fermion Resonances}

Here we generalize our previous results where resonances of fermions for $s=1$ were analyzed \cite{Landim:2011ki}. The study of zero mode localization of this field has been performed extensively in 
the last years \cite{Liu:2009ve,Zhao:2009ja,Liang:2009zzf,Zhao:2010mk,Zhao:2011hg,Li:2010dy,
Castro:2010au,Correa:2010zg,Castro:2010uj,Chumbes:2010xg,Castro:2011pp}. Here we look for resonant modes of this field. We pointed out before that this depends 
strongly on the form of the potential and on the model used. Therefore, when we consider the process of deformation, the potential gets modifications that will alter the spectrum of resonances and the 
phenomenology of the problem. Here we must consider a model similar to that found  in \cite{Kehagias:2000au,Li:2010dy,Landim:2011ki}. 
We must give only a short review on how to arrive to the Schr\"odinger equation and of the associated potential. The action considered is slightly modified to include the
dilaton coupling. Consider the action 
\begin{eqnarray}
 S_{1/2} &=& \int d^{5}x\sqrt{-g}e^{-\lambda\pi}\left[\overline{\Psi}\Gamma^{M}D_{M}\Psi
    -\eta\overline{\Psi}F(\phi)\Psi\right],
    \label{fermion field action}
\end{eqnarray}
where $D_M=\partial_M+\omega_M$, $\omega_M$ being the spin connection. The $\Gamma^M=e_{N}^{~~M}\gamma^N$ are the Dirac matrices in the five dimensional curved space-time 
and $e_{N}^{~M}$ are the vielbeins: $e_{A}^{~~M}e_{B}^{~~N}\eta^{AB}=g^{MN}$. After some manipulations with spin connection we get the equations of motion

\begin{eqnarray}
\left[\gamma^{\mu}\partial_{\mu}+e^{A_s-B_s}\gamma^{5}\left(\partial_{y}+2\partial_{y}A_s\right)-\eta
e^{A_s}F(\phi)\right]\Psi=0\,,
 \label{Dirac}
\end{eqnarray}
where $\gamma^{\mu}\partial_{\mu}$ is the four-dimensional Dirac
operator in the brane. It is interesting to note here that the dilaton coupling just modifies the equation of motion throughout the $B_s$ factor of the metric. The value of $\lambda$ 
will be irrelevant to the resonance spectrum analysis. Now, just as in the gravity case we perform the decomposition
\begin{eqnarray}
\Psi(x,y)=e^{-2A_s}\left(\sum_{n}\psi_{Ln}(x)f_{Ln}(y)+\sum_{n}\psi_{Rn}(x)f_{Rn}(y)\right).
\label{fermion decomposition}
\end{eqnarray}

In the above we defined $\gamma^{5}\psi_{Ln}(x)=-\psi_{Ln}(x)$ and
$\gamma^{5}\psi_{Rn}(x)=\psi_{Rn}(x)$ as the left and right-handed fermion fields respectively. Just as pointed in the second section we obtain 
from the above the equation for a massive field in four dimension 
$$\gamma^{\mu}\partial_{\mu}\psi_{Ln}(x)=m_{n}\psi_{Rn}(x)$$
 and
$$\gamma^{\mu}\partial_{\mu}\psi_{Rn}(x)=m_{n}\psi_{Ln}(x),$$
with mass defined by
\begin{eqnarray}
 \left[\partial_{y} +{\eta} e^{B_s}F(\phi)\right]
  f_{Ln}(y) &=& \,\,\,\,\,m_{n}e^{B_s-A_s}f_{Rn}(y)\,,\label{extradimension1} \\
 \left[\partial_{z}-\eta e^{B_s}F(\phi)\right]
  f_{Rn}(y) &=& -m_{n}e^{B_s-A_s}f_{Ln}(y)\,. \label{extradimension2}
\end{eqnarray}

From the above equations we can obtain the Schr\"odinger equation for each chirality

\begin{eqnarray}
 \left[-\partial_{z}^{2}+U_{L}(z)\right]f_{L} &=& m^{2}f_{L} \,,
    \\ \label{ScheqLeft}
 \left[-\partial_{z}^{2}+U_{R}(z)\right]f_{R} &=& m^{2}f_{R} \,,
       \label{ScheqRight}
\end{eqnarray}
with potential given by
\begin{eqnarray}
U_{L}(z)=e^{2A_s-B_s}\left(\eta^{2} e^{B_s}F^{2}(\phi)-\eta
F'(\phi)-\eta
(A'_s)F(\phi)\right)\,,  \label{VzL}  \\
U_{R}(z)=e^{2A_s-B_s}\left(\eta^{2} e^{B_s}F^{2}(\phi)+\eta
F'(\phi)+\eta (A'_s)F(\phi)\right)\,,
\label{fermioneq}
\end{eqnarray}
with $dz/dy=e^{B_s-A_s}$.

As before, for the case without the dilaton, we just fix $B_s=0$ to obtain the respective potentials
\begin{eqnarray}
U_{L}(z)=e^{2A_s}\left(\eta^{2}F^{2}(\phi)-\eta
F'(\phi)-\eta
(A'_s)F(\phi)\right)\,,   \\
U_{R}(z)=e^{2A_s}\left(\eta^{2} F^{2}(\phi)+\eta
F'(\phi)+\eta (A'_s)F(\phi)\right)\,.
\label{fermioneq}
\end{eqnarray}

We show in Figs. \ref{fig:s3-Fermion-eta10} and \ref{fig:s5-Fermion-eta10} the graphic of the potentials with and without the dilaton with $F(\phi)=\phi$, for $s=3$ and $s=5$ respectively.
It is important to note that, different than the $s=1$ case, both profiles are similar to a double barrier and we should expect resonances. We have previously commented \cite{Landim:2011ki} that the 
value of $\eta$ can induce the presence of resonances. Here we discover that the value of $s$ can also provide the increasing the number of resonances. We use only the value $\eta=10$ but if we compare
the Figs. \ref{fig:logts3-Fermion-eta10} and \ref{fig:logts5-Fermion-eta10} we see that the case $s=5$ has more peaks than $s=3$ regardless to the dilaton coupling. Therefore it seems that although the dilaton coupling can induce the resonances, the value of $s$ has a stronger role on this. We should stress that in comparison with the $s=1$ case we have a richer structure of resonances, as mentioned in the introduction.

\FIGURE{
\centerline{\psfig{figure=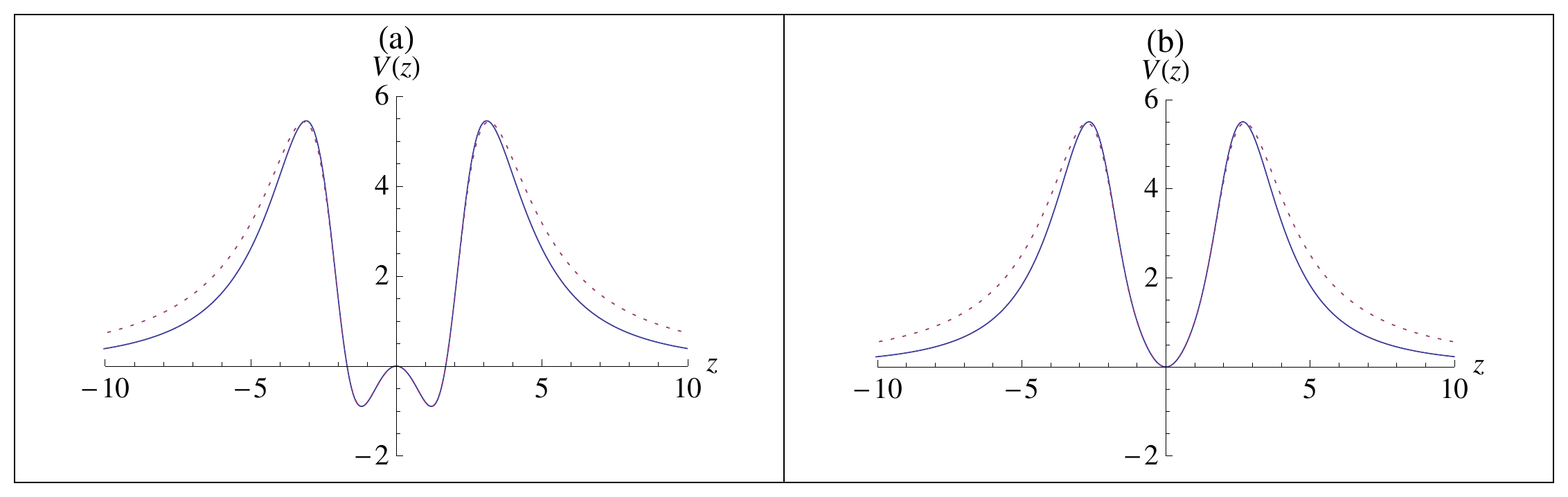,angle=0,height=4cm}}
\caption{Potential of the Schr\"odinger like equation of fermion for $s=3$ with $\eta=10$. (a) Left without dilaton (dotted) and with dilaton (lined) and (b) Right without dilaton (dotted) and with 
dilaton (lined).\label{fig:s3-Fermion-eta10}}
}

\FIGURE{
\centerline{\psfig{figure=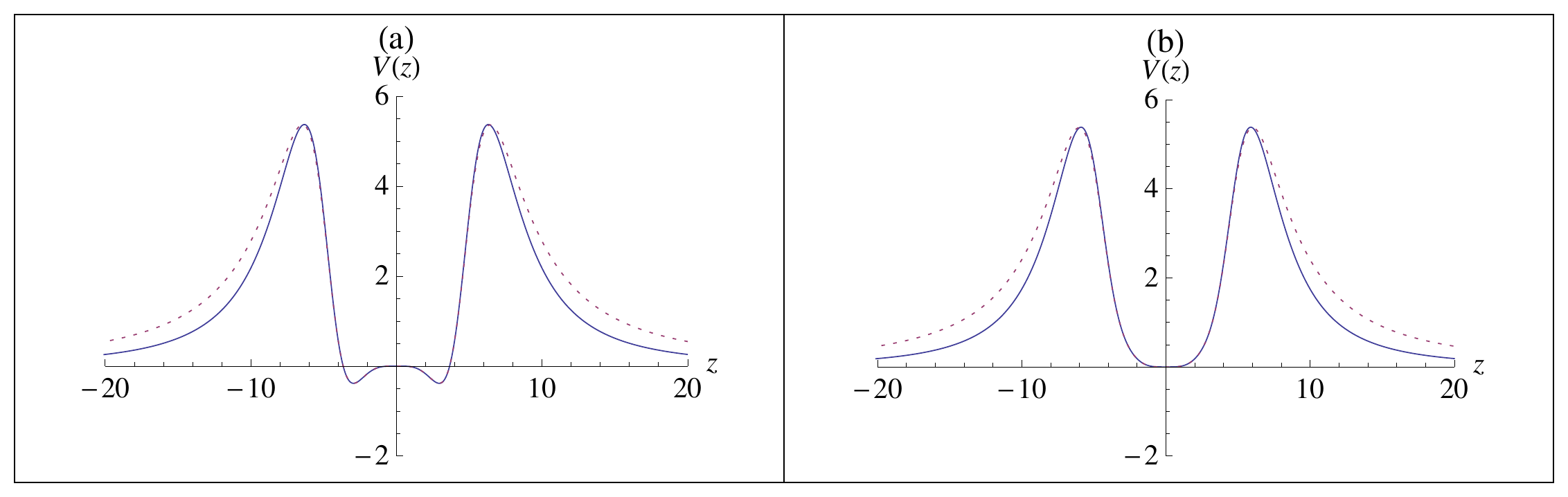,angle=0,height=4cm}}
\caption{Potential of the Schr\"odinger like equation of fermion for $s=5$ with $\eta=10$ . (a) Left without dilaton (dotted) and with dilaton (lined) and (b) Right without dilaton (dotted) and with 
dilaton (lined).\label{fig:s5-Fermion-eta10}}
}

\FIGURE{
\centerline{\psfig{figure=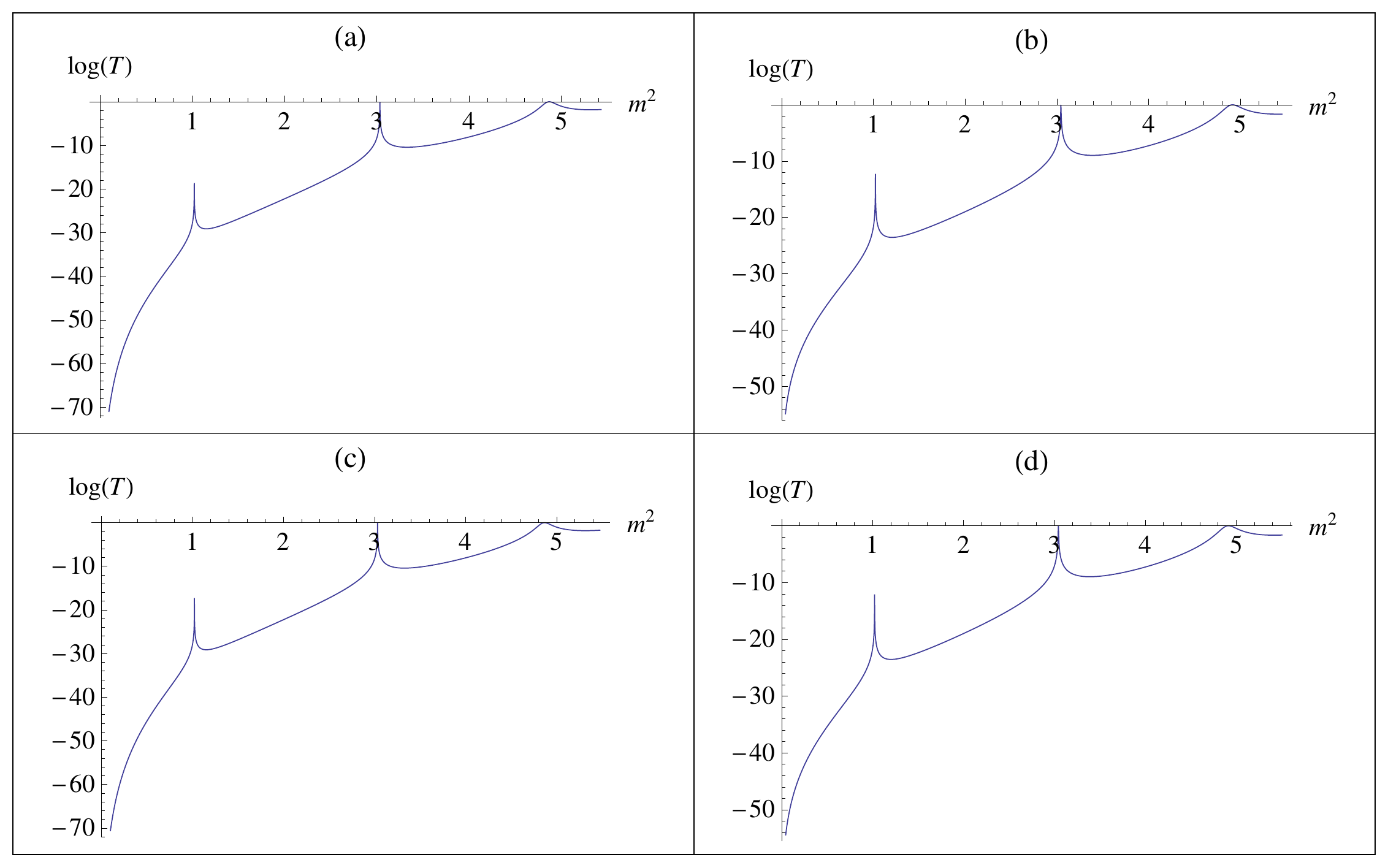,angle=0,height=8cm}}
\caption{Logarithm of the transmission coefficient of fermion for $s=3$ with $\eta=10$. (a) Left without dilaton, (b) Left with dilaton, (c) Right without dilaton and (d) Right with dilaton.
\label{fig:logts3-Fermion-eta10}}
}

\FIGURE{
\centerline{\psfig{figure=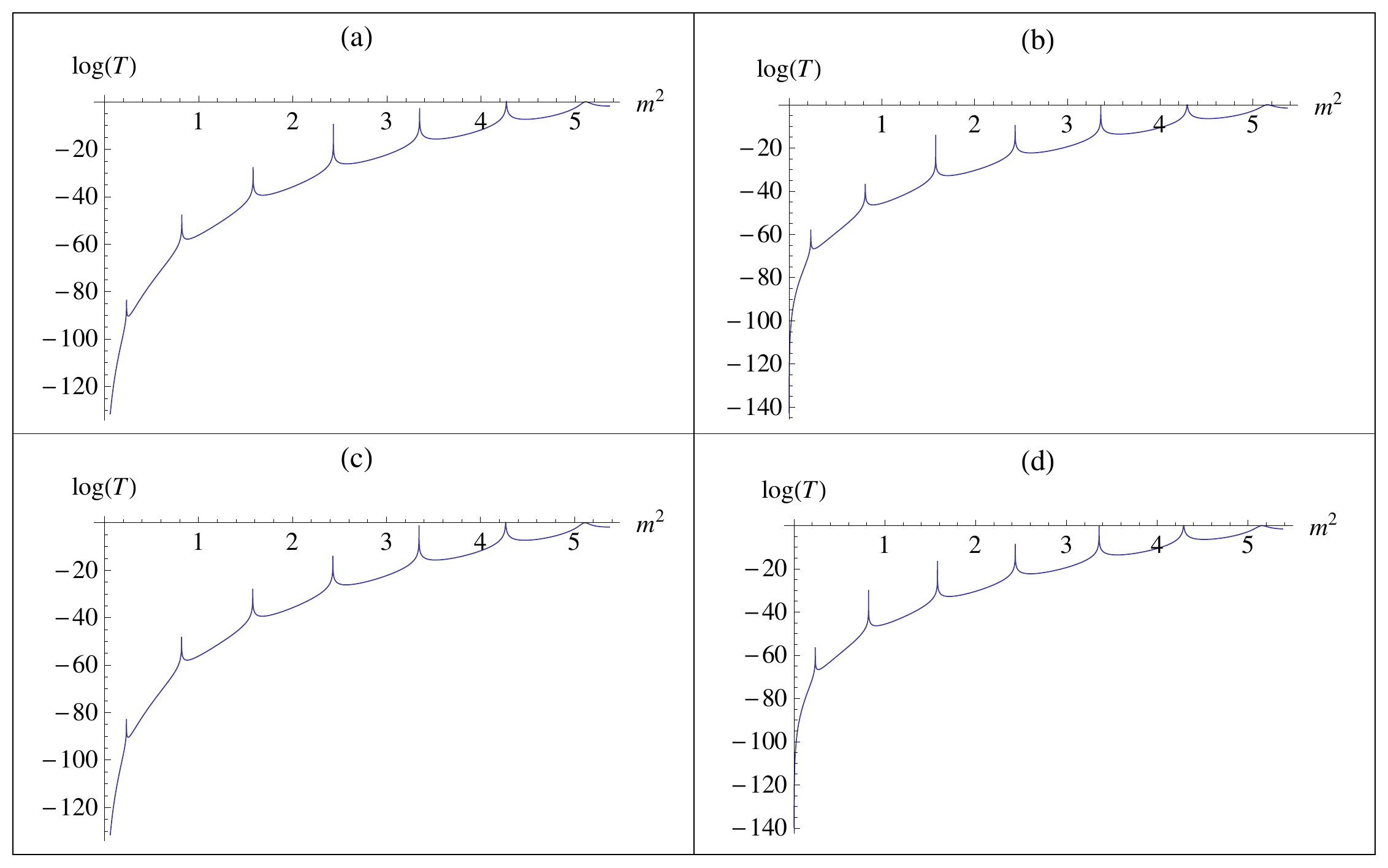,angle=0,height=8cm}}
\caption{Logarithm of the transmission coefficient of fermion for $s=5$ with $\eta=10$. (a) Left without dilaton, (b) Left with dilaton, (c) Right without dilaton and (d) Right with dilaton.
\label{fig:logts5-Fermion-eta10}}
}

\section{Conclusions}

In this manuscript we have studied the issue of resonances in a deformed Randall-Sundrum-like model. For this we used a tool from quantum mechanics and condensed matter physics: the Transfer Matrix method. 
In all Randall-Sundrum-like models the issue of localization is studied through an associated one dimensional Schr\"odinger equation with an associated potential. The general prescription imposes that 
in the limit $y\to \infty$ the Randall-Sundrum warp factor should be recovered. With this the potential of the associated Schr\"odinger equation has the general behavior of going to zero at infinity. This 
justifies the idea that we must consider plane waves at infinity that will interact with the membrane. This approach has been put forward successfully in Ref. \cite{Landim:2011ki} where the resonant peaks of many bulk fields in a kink-like membrane were study. The richness of this model can be widened if we consider another sort of topological defects. This can be obtained in a simple way by a deformation procedure. Due to this deformation all potentials of the associated Schr\"odinger equation are changed and as a consequence the spectrum of resonances is also modified. The interesting point here is that, different from the undeformed case we have a very rich structure of resonances. For almost all cases analyzed here we have found more peaks than the cases studied previously. With this method we have computed resonances of gravity, fermions and gauge fields. 

The results obtained here corroborate our claim that, beyond the coupling with the dilaton, the deformed defect can give us a richer structure of resonances. The only case in which this is not evidenced is 
the gravity one. We see in Figs. \ref{fig:grav-logts3} and \ref{fig:grav-logts5} that we only have one resonance for any case considered. When we studied the case of form fields, namely the scalar, vector and 
Kalb-Ramond fields that became more apparent. If we compare the number of resonances for the scalar field with $s=1$\cite{Landim:2011ki} and with $s=3,5$ in Figs. \ref{fig:logts3q0} (a) and \ref{fig:logts5q0} (a) we 
found $6$,$15$ and $20$ respectively. The vector and Kalb-Ramond fields have similar behavior. The analyzes of the fermion resonances also corroborate these ideas. For this case we found in 
Figs. \ref{fig:logts3-Fermion-eta10} and \ref{fig:logts5-Fermion-eta10} many peaks of resonances that rises when we increase the value of $s$. For example, if we 
compare Figs. \ref{fig:logts3-Fermion-eta10} (d) and \ref{fig:logts5-Fermion-eta10} (d) for the right fermion with the dilaton we see three peaks for $s=3$ and seven 
peaks for $s=5$. It is important mentioning that for these phenomenological models, many massive modes are expected to interact with the membrane.

Furthermore, there are some applications of the models studied here related to AdS/CFT correspondence: there are evidences supporting the idea of domain wall/QFT, i.e., a 
correspondence between gauged supergravities and quantum field theories in domain walls. Important applications are related to the study of quark-gluon plasma via gravity-duals. As a perspective, we will try to adapt this numerical method to attack such kind of problem.

\section*{Acknowledgment}

We would like to thank the Laborat\'orio de \'Oleos pesados. We also acknowledge the 
financial support provided by Funda\c c\~ao Cearense de Apoio ao Desenvolvimento Cient\'\i fico e Tecnol\'ogico (FUNCAP), the Conselho Nacional de Desenvolvimento Cient\'\i fico e Tecnol\'ogico (CNPq) and FUNCAP/CNPq/PRONEX.

This paper is dedicated to the memory of my wife  Isabel Mara (R. R. Landim).

\end{document}